\documentclass[aps,prx,reprint,twocolumn,nofootinbib,amsmath,amssymb,amsfonts]{revtex4-1}

\AtBeginDocument{\usepackage{booktabs}}               
\makeatletter
\g@addto@macro\bfseries{\boldmath}
\makeatother

\usepackage[varg]{txfonts}
\usepackage[T1]{fontenc}
\usepackage[utf8]{inputenc}

\usepackage{hyperref}
\usepackage{amsmath}
\usepackage{color}
\usepackage{graphicx}
\usepackage[percent]{overpic}
\usepackage{mathrsfs}
\usepackage{bm}
\usepackage{braket}
\usepackage[nolist,nohyperlinks]{acronym}

\newcommand{\tsup}[1]{\textsuperscript{#1}}
\newcommand{\tsub}[1]{\textsubscript{#1}}

\newcommand{\subref}[2]{\ref{#1}\hyperref[#1]{#2}}

\newcommand{\avg}[1]{\braket{#1}}

\newcommand{\h}[1]{{#1}^{\dagger}}
 
\newcommand{\cc}[1]{{#1}^{*}}
\newcommand{\cb}[1]{\bar{#1}}

\newcommand{\hc}{{\rm h.c.}}
\newenvironment{subalign}{\subequations\align}{\endalign\endsubequations}

\renewcommand{\vec}[1]{\boldsymbol{#1}}
\newcommand{\mat}[1]{\vec{#1}}
\newcommand{\trp}[1]{{#1}^{\intercal}}
\newcommand{\vhat}[1]{\vec{\hat{#1}}}

\newcommand{\abo}[2]{#1\tsub{2}#2\tsub{2}O\tsub{7}}

\newcommand{\yto}{\abo{Yb}{Ti}}
\newcommand{\yso}{\abo{Yb}{Sn}}
\newcommand{\ygo}{\abo{Yb}{Ge}}
\newcommand{\rth}[1]{{#1}\tsup{3+}}

\newcommand{\mueV}{\ \mu{\rm eV}}
\newcommand{\meV}{\ {\rm meV}}
\newcommand{\T}{\ {\rm T}}
\newcommand{\mT}{\ {\rm mT}}

\newcommand{\mK}{\ {\rm mK}}

\definecolor{cred}{RGB}{228,26,28}
\definecolor{cblue}{RGB}{55,126,184}
\definecolor{cdblue}{RGB}{40,96,139}
\definecolor{clblue}{RGB}{205,223,237}
\definecolor{cgreen}{RGB}{77,175,74}
\definecolor{cgray}{RGB}{150,150,150}
\definecolor{clgray}{RGB}{200,200,200}
\definecolor{cpurple}{RGB}{152,78,163}
\definecolor{corange}{RGB}{255,127,0}
\definecolor{cgold}{RGB}{230,171,2}

\definecolor{cL}{RGB}{180,60,60}
\definecolor{cLL}{RGB}{255,150,150}
\definecolor{cG}{RGB}{204,204,204}

\hypersetup{colorlinks=true,linkcolor=cL,citecolor=cL,urlcolor=cL} 

\newcommand{\citetalt}[1]{{\hypersetup{citecolor=cLL}\citet{#1}\hypersetup{citecolor=cL}}}     

\newcommand{\parThompson}{\citet{thompson2017}}

\newacro{DM}[DM]{{Dzyaloshinskii-Moriya}}
\newacro{AIAO}[AIAO]{all-in/all-out}
\newacro{PC}[PC]{Palmer-Chalker}
\newacro{SFM}[SFM]{splayed ferromagnet}
\newacro{LSWT}[LSWT]{linear spin-wave theory}
\newacro{NLSWT}[NLSWT]{non-linear spin-wave theory}
  
\begin{document}

\title{Magnon interactions in the frustrated pyrochlore ferromagnet \yto{}}
\author{Jeffrey G. Rau} 
\author{Roderich Moessner} 
\author{Paul A. McClarty}
\affiliation{Max-Planck-Institut f\"ur Physik komplexer Systeme, 01187 Dresden, Germany}

\begin{abstract}
  The frustrated rare-earth pyrochlore \yto{} is remarkable among magnetic materials: despite a ferromagnetically ordered ground state it exhibits a broad, nearly gapless, continuum of excitations. This broad continuum connects smoothly to the sharp one-magnon excitations expected, and indeed observed, at high magnetic fields, raising the question: how does this picture of sharp magnons break down as the field is lowered?  In this paper, we consider the effects of magnon interactions in \yto{}, showing that their inclusion greatly extends the reach of spin-wave theory. First, we show that magnon interactions shift the phase boundary between the (splayed) ferromagnet (SFM) and the antiferromagnetic $\Gamma_5$ phase so that \yto{} lies very close to it. Next, we show how the high-field limit connects to lower fields; this includes corrections to the critical fields for the $[111]$ and $[1\bar{1}0]$ directions, bringing them closer to the observed experimental values, as well as accounting for the departures from linear spin wave theory that appear in $[001]$ applied fields below $3\T$ [Thompson \emph{et al.}, Phys. Rev. Lett. \textbf{119}, 057203 (2017)]. Turning to low-fields, though the extent of the experimentally observed broadening is not quite reproduced, we find a rough correspondence between non-linear spin-wave theory and inelastic neutron scattering data on both a single-crystal sample, as well as on a powder sample [Pe\c{c}anha-Antonio \emph{et al.}, Phys. Rev. B, \textbf{96}, 214415 (2017)]. We conclude with an outlook on implications for future experimental and theoretical work on \yto{} and related materials, highlighting the importance of proximity to the splayed ferromagnet-$\Gamma_5$ phase boundary and its potential role in intrinsic or extrinsic explanations of the low-field physics of \yto{}.
\end{abstract}

\date{\today}

\maketitle

\section{Introduction}
\label{sec:Introduction}

Of the hundreds of known three dimensional and (anisotropic) two dimensional magnetic materials, the overwhelming majority order magnetically at low temperatures and have magnetic excitations that may be understood as the quasiparticle modes corresponding to harmonic fluctuations of the underlying magnetic structure. One of the grand endeavours of condensed matter physics has been to identify materials where quantum fluctuations cause departures from this canonical picture. A particularly fruitful strategy to find such systems has been to focus on materials where the magnetic interactions are frustrated~\cite{Lacroix}, such as through geometrical frustration on lattices of corner-sharing triangles or through frustration arising from intrinsically anisotropic exchange couplings.

A noteworthy example is found in \yto{}: a rare earth magnet with effective spin one-half ytterbium ions on the pyrochlore lattice of corner-sharing tetrahedra~\cite{Gardner2010,rau2019,HallasReview,Blote,Hodges2001a,Hodges2001,Hodges2002,Yasui2003,Bonville2004,Gardner2004,Malkin2004,Cao2009,Cao2009a,Ross2009,Thompson2011a,Thompson2011,Onoda2011a,Onoda2011,Yaouanc2011,Ross2011b,Chang2012,Hayre2012,Applegate2012a,ross2011,Ross2012a,DOrtenzio2013,Chang2014,Lhotel2014,pan2014low,Jaubert2015,Robert2015,gaudet2015,Pan2015,Bhattacharjee2016,Tokiwa2016,gaudetgapless2016,yaouanc2016novel,yan2017,VPA2017,kermarrec2017ground,thompson2017,Arpino2017,Mostaed2017,li2018low,shafieizadeh2018superdislocations}. The magnetic ground state of \yto{} has been directly found to be ferromagnetic with a net spontaneous moment along a cubic lattice direction~\cite{Yasui2003,Chang2012,yaouanc2016novel,kermarrec2017ground}, further supported by the existence of symmetry-breaking phase transitions in $[1\bar{1}0]$~\cite{ross2011} and $[111]$ magnetic fields~\cite{scheie2017} and by the absence of such a transition in a $[001]$ magnetic field~\cite{thompson2017}. The finite temperature transition into the low temperature magnetically ordered state is known to be sensitive to the details of sample synthesis~\cite{Blote,Hodges2001a,Bonville2004,Yasui2003,Yaouanc2011,Ross2012a,Lhotel2014,gaudet2015,gaudetgapless2016,Mostaed2017} but samples have been made in which a sharp transition is observed in the range $200-260$ mK~\cite{Chang2012,thompson2017,scheie2017}.  

In striking contrast to conventional magnets, the magnetic excitations in \yto{} in {\it zero} applied magnetic field are dominated by an apparent broad continuum of intensity. This behavior is reminiscent of the low-energy excitations observed in the (disordered) kagome antiferromagnet Herbertsmithite~\cite{han2012fractionalized} and the higher-energy features seen in the (ordered) honeycomb antiferromagnet $\alpha$-RuCl\tsub{3}, which both exhibit anomalously broad excitation spectra~\cite{banerjee2016proximate,banerjee2017neutron,banerjee2018excitations}. However, unlike these materials, in \yto{} the exchange scale is small and the magnetic moments relatively large, so the magnetic ground state is tunable using easily accessible laboratory magnetic fields. This allows access to a high field regime where sharp one-magnon excitations are observed~\cite{ross2011,thompson2017}. 

In a single crystal with a sharp transition at $\sim 214 \mK$, the magnetic excitations have been studied in magnetic fields along $[001]$ from high-fields ($9\T$) down to zero field~\cite{thompson2017}. This experiment revealed the presence of sharp one-magnon excitations, well separated from the two-magnon excitations, that are well-described by \ac{LSWT} down to magnetic fields of about $3\T$.  As the magnetic field was lowered further, the one-magnon dispersions were significantly renormalized from the \ac{LSWT} predictions and level repulsion from the two magnon continuum was inferred. At low fields, as the two magnon continuum begins to overlap with the one magnon states one finds pronounced broadening effects. At zero field, the scattering intensity over large areas of momentum space was found to be a broad continuum within the resolution of the instrument.

These experimental results present two clear puzzles. Foremost is the question of how the magnetically ordered ground state and the continuum of excitations can be understood in a unified way that also smoothly recovers the conventional behavior observed at high-fields.  Secondly, one would like to understand the origin of the sample dependence of the magnetic transition and whether this is related to the exotic zero field behavior. In this paper, we tackle the first of these questions by looking at the leading corrections to spin wave theory beyond the non-interacting theory.

The observed departures of the inelastic intensity from \ac{LSWT} at intermediate fields are of a nature that one might expect to be described by extending \ac{LSWT} to include the effects of magnon interactions. It is well understood that, generically, these interactions include terms that violate magnon number conservation - including three-magnon couplings~\cite{zhitomirsky2013colloquium,chubukov1994large,chernyshev2009triangular,mourigal2013dynamical}. These couplings induce spontaneous decay processes that mix the one-magnon and the two-magnon sectors. Such decays are kinematically forbidden when there is a separation in energy between the one- and two-magnon states (as is the case at high fields), only becoming operative at low fields when there is overlap between the two in energy-momentum space. Their effects have been extensively studied theoretically in various models of low-dimensional frustrated magnets where quantum fluctuations are {\it a priori} expected to be significant~\cite{zhitomirsky2013colloquium,chubukov1994large,chernyshev2009triangular,mourigal2013dynamical,plumb2016,hwang2016,PhysRevB.98.155102,verresen2018strong}. Experimentally, such effects have also been observed in several different magnetic materials using inelastic neutron scattering~\cite{Masuda2010,Oh2013,robinson2014quasiparticle,dalla2015fractional,sandilands2015,hong2017field}. In \yto{}, the experimental evidence is clear that the onset of the broadening of one-magnon intensity coincides with the merging of the two-magnon continuum with the higher energy one-magnon branches~\cite{thompson2017}. In addition, the observed broadening proceeds to lower energies as the field is further lowered and the multi-magnon states begin to significantly overlap the remaining sharp one-magnon modes.

As much as one might hope that \ac{NLSWT} can capture the intermediate field inelastic properties of \yto, one also naively expects it to fail to describe the zero field features. This is because the computed magnon spectrum about the experimentally determined ferromagnetic state is gapped; the lowest energy multi-magnon states then lie at a finite threshold energy above the lowest energy one-magnon states, freezing out any potential decay channels for the lowest energy modes. In contrast, the experimentally observed spectrum appears to be broadened down to the lowest accessible energies suggesting that any gap is relatively small, if present at all. In fact, even in the extreme case where the spectrum is gapless, the decay matrix elements and density of decay states may be such that one-magnon intensity remains relatively sharp despite the presence of kinematically allowed one-magnon decay down to zero energy.

While the above remarks about the role of magnon interactions are largely model-independent, it is desirable to have a concrete calculation of their effects in \yto{} to establish the \emph{quantitative} extent to which they capture the experimental data coming down from high fields. Speculations about the origin and nature of the observed continuum are moot in the absence of such a calculation so, in this paper, we directly address this issue. That this question can be meaningfully addressed in a quantitative manner rests on detailed fits of the sharp one-magnon dispersion measured in a regime of high fields where interaction effects can be neglected and \ac{LSWT} is adequate to describe the physics~\cite{ross2011,thompson2017}. 

Due to the strong spin-orbit coupling of the $4f^{13}$ states of \rth{Yb}, the symmetry-allowed exchanges can be highly anisotropic, as is indeed found in the parameters that have been determined empirically.
We base our analysis on the parameter set of Ref.~\cite{thompson2017}, determined by comparision to the spectrum in high magnetic fields. These exchange parameters place \yto{} close to the (semi-classical) phase boundary between the \ac{SFM} of \yto{} and an antiferromagnetic $\Gamma_5$ phase. It has been suggested that proximity to this phase boundary underlies the exotic properties of \yto{}~\cite{Jaubert2015,Robert2015}. This picture is consistent with the fact that hydrostatic pressure tunes samples that show no signs of long-range order into the \ac{SFM} state~\cite{kermarrec2017ground}. 

The paper is organized as follows. First, in Sec.~\ref{sec:model-methods}, we describe the anisotropic exchange model for \yto{} considered in the remainder of the paper (Sec.~\ref{sec:model}) and then outline our approach to \ac{NLSWT} (Sec.~\ref{sec:swt}).

In Sec.~\ref{sec:competing-phases}, we consider \ac{NLSWT} corrections to static properties, in particular the \ac{SFM}-$\Gamma_5$ phase transition and critical magnetic fields. In Sec.~\ref{sec:phase}, we study the zero field phase boundary using \ac{NLSWT}, finding that the \ac{SFM} phase becomes unstable near the boundary with the $\Gamma_5$ phase. We show that the parameters of Ref.~\cite{thompson2017}, which are close to the classical phase boundary, sit almost exactly on top of this \ac{NLSWT} instability line. This is our first main result: that the effects of magnon interactions drive \yto{} even closer to the \ac{SFM}-$\Gamma_5$ phase boundary. In Sec.~\ref{sec:critical-fields}, we study  the phase transitions in applied magnetic fields along $[1\bar{1}0]$ and $[111]$ directions. Taking into account corrections from \ac{NLSWT}, we find the critical fields depart from their classical values, which crucially bring them closer to what has been observed experimentally. This is our second main result.

We now turn to the key question: how does \ac{LSWT} breakdown as the fields are lowered? These results are presented in Sec.~\ref{sec:DSF}, where we find that the regime of (semi-)quantitative reliability of spin-wave theory is significantly enhanced by the inclusion of magnon interactions. This is supported by a detailed study of the dynamical structure factor computed in \ac{NLSWT} for three high symmetry field directions  $[001]$ (Sec.~\ref{sec:cuts-001}), $[1\bar{1}0]$ (Sec.~\ref{sec:cuts-1m10})  and $[111]$ (Sec.~\ref{sec:cuts-111}). Next, in Sec.~\ref{sec:direct-comparison}, we make a direct comparison with the existing experimental data of \citet{thompson2017} for $[001]$ magnetic fields. For fields larger than $\sim 3\T$, where \ac{LSWT} has been applied successfully, we confirm that departures from \ac{LSWT} are small. For lower fields, we find that \ac{NLSWT} captures well the renormalizations of the one-magnon modes down to fields of about $|\vec{B}| \sim 0.5\T$. For $[1\bar{1}0]$ fields we confirm the validity of \ac{LSWT} for $|\vec{B}| \gtrsim 2\T$, as had been assumed in previous works~\cite{ross2011}. For $[111]$ fields, we find significant departures from \ac{LSWT} for fields as high as $|\vec{B}| \sim 3\T$. In Sec.~\ref{sec:powder}, we also provide a comparison to the high-resolution powder-averaged zero field inelastic data of \citet{VPA2017}, finding rough qualitative agreement.

In Sec.~\ref{sec:disc}, we summarize and discuss our results and their implications for \yto{}. In Sec.~\ref{sec:outlook}, we discuss possible origins of the anomalous zero-field behavior, including intrinsic and extrinsic scenarios, as well as the importance of proximity to the \ac{SFM}-$\Gamma_5$ phase boundary.


\section{Model and methods}
\label{sec:model-methods}

\begin{figure}[htp]
    \centering
    \begin{overpic}[width=0.8\columnwidth]
    {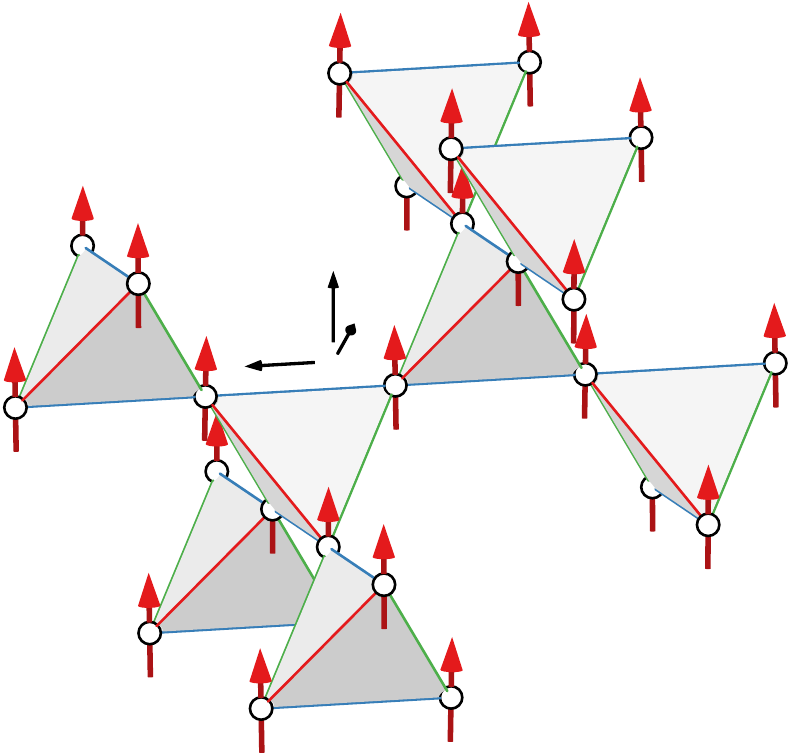}
    \put(43,13){\textcolor{cred}{$x$}}
    \put(46,8){\textcolor{cblue}{$z$}}
    \put(50,13){\textcolor{cgreen}{$y$}}
    \put(34,57){$\vhat{B}$}
    \put(38,63){\scriptsize{[001]}}
    \put(30,51){\scriptsize{[1$\bar{1}$0]}}
    \put(44,56){\scriptsize{[111]}}
    \put(12,15){$\vec{\mu}_r$}
    \end{overpic}
    \caption{\label{fig:lattice}
    Pyrochlore lattice of \rth{Yb} ions in \yto{}. The three symmetry related nearest-neighbor bond types ($x$, $y$ and $z$) are labelled, along with the three magnetic field directions we consider in this work. Also shown are the directions of the magnetic moments, $\vec{\mu}_r$, of the \rth{Yb} ions in the \acf{SFM} state expected (classically) at zero field for the parameters of \citet{thompson2017}.}
\end{figure}
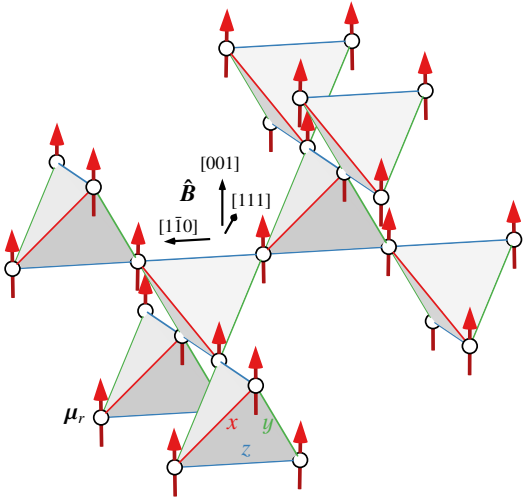

We first describe the appropriate model for the magnetic ions in \yto{} and then outline our approach to \ac{NLSWT}. Our treatment of magnon interactions is described briefly and we refer the reader to standard treatments for further details~\cite{chubukov1994large,chernyshev2009triangular,zhitomirsky2013colloquium}.
\subsection{Anisotropic exchange model}
\label{sec:model}
We consider the general nearest-neighbor model for effective spin-$1/2$ Kramers doublets on the pyrochlore lattice~\cite{PhysRevB.78.094418,mcclarty2009energetic,Onoda2011,Onoda2011a,ross2011,PhysRevB.98.054408}. One can restrict to the crystal field ground doublet in \yto{}, as the gap to the next crystal field doublet~\cite{gaudet2015} is much larger than the exchange scale and the temperature. The appropriate effective model is then
\begin{align}
\label{eq:model} 
    H &\equiv \sum_{\avg{rr'}} \left[  J_{zz} S^z_r S^z_{r'} -
J_{\pm}\left(S^+_r S^-_{r'}+S^-_r S^+_{r'}\right)+ \right. \nonumber \\ &
\left. J_{\pm\pm} \left(\gamma_{ij} S^+_r S^+_{r'}+\hc \right)+ 
J_{z\pm}\left( \zeta_{ij} \left[ S^z_r S^+_{r'}+ S^+_r S^z_{r'} \right]+ \hc \right)\right] \nonumber \\  & - \sum_r \vec{B} \cdot \vec{\mu}_r,
\end{align}
where the $\gamma_{rr'}$ and $\zeta_{rr'}$ are bond dependent phases and
\begin{equation}
  \label{eq:moment}
    \vec{\mu}_r \equiv  -\mu_B\left[
    g_{\pm} \left(S^x_r \vhat{x}_r +
      S^y_r \vhat{y}_r\right) +g_{z} S^z_r \vhat{z}_r \right].
\end{equation}
The relevant phases factors $\gamma_{ij}$ and $\zeta_{ij}$ for the three nearest-neighbor bond types (labelled $x$, $y$ and $z$, shown in Fig.~\ref{fig:lattice}) are
\begin{align}
  \gamma^{}_{x} = -\cc{\zeta}_x &= 1, &
  \gamma^{}_{y} = -\cc{\zeta}_y &= \omega, &
  \gamma^{}_{z} = -\cc{\zeta}_z &= \omega^2,
\end{align}
where $\omega = e^{2\pi i/3}$ and the local axes, $(\vhat{x}_r,\vhat{y}_r,\vhat{z}_r)$, follow the convention of \citet{savary2012} (see App.~\ref{app:conventions}).  

We primarily consider the parameters obtained in \citet{thompson2017}
\begin{align}
  J_{zz} &= +0.026 \meV,   & J_{\pm} &=+0.074 \meV, \nonumber \\ \nonumber  J_{\pm\pm} &= +0.048 \meV, &  J_{z\pm} &= -0.159 \meV, \\
  g_{z} & =2.14,  & g_{\pm} &=4.17,
\end{align}
where the Ising coupling is small, and $J_{z\pm}$ is the dominant exchange. These were obtained through fits to  the magnetic excitations in high-fields in both $[1\bar{1}0]$ and $[001]$ directions~\cite{thompson2017} and are, further, consistent with the terahertz measurements of Ref.~\cite{pan2014low}. Earlier parametrizations, obtained using inelastic neutron scattering data at high field using a single field orientation, \citet{ross2011} and \citet{Robert2015}, also lie in the classical ferromagnetic phase corresponding to the experimentally determined ground state. These differ, however, in detail, with the parameters of \citet{ross2011} being somewhat further from the classical phase boundary between the \ac{SFM} and the $\Gamma_5$ states, than those of \citet{thompson2017} or \citet{Robert2015}.

\subsection{Spin wave theory}
\label{sec:swt}

To make our discussion somewhat general, we write the nearest-neighbor exchange model [Eq.~(\ref{eq:model})] in the form
\begin{equation}
  H \equiv \frac{1}{2} \sum_{\vec{r}\vec{r}'} \sum_{\alpha\alpha'} \trp{\vec{S}}_{\vec{r}\alpha}\mat{J}_{\vec{r}-\vec{r}',\alpha\alpha'} \vec{S}_{\vec{r}'\alpha'},
\end{equation}
where $\vec{S}_{\vec{r}\alpha}$ is a spin-$S$ operator located in unit cell $\vec{r}$ with sublattice index $\alpha$.  The spin operators can expressed in terms of Holstein-Primakoff bosons~\cite{holstein1940,kittel1963quantum,auerbach1998interacting} as
\begin{align}
  \vec{S}_{\vec{r}\alpha}  \equiv &
  \sqrt{S}\left[
    \left(1-\frac{n_{\vec{r}\alpha}}{2S}\right)^{1/2} {a}^{}_{\vec{r}\alpha}  \vhat{e}_{\alpha,-}+
    \h{a}_{\vec{r}\alpha}  \left(1-\frac{n_{\vec{r}\alpha}}{2S}\right)^{1/2}
    \vhat{e}_{\alpha,+}\right] \nonumber \\
  & +\left(S-n_{\vec{r}\alpha}\right)\vhat{e}_{\alpha,0},
\end{align}
where $n_{\vec{r}\alpha} \equiv \h{a}_{\vec{r}\alpha} {a}^{}_{\vec{r}\alpha}$ and vectors $\vhat{e}_{\alpha,\pm}$, $\vhat{e}_{\alpha,0}$ define a local frame of reference with $\vhat{e}_{\alpha,0}$ being the classical ordering direction. It is useful to write the exchange matrix in this frame
\begin{equation}
  \label{eq:lswt:exchange}
  \mathcal{J}^{\mu\mu'}_{\vec{\delta},\alpha\alpha'} \equiv \trp{\vhat{e}}_{\alpha,\mu} \mat{J}_{\vec{\delta},\alpha\alpha'}{\vhat{e}}^{}_{\alpha',\mu'},
\end{equation}
where $\mu,\mu' = 0,\pm$.

Expanding in powers of $1/S$ then yields a semi-classical expansion
about the ordered state defined by $\vhat{e}_{\alpha,0}$.
At $O(1/S^2)$, this can be written as a sum of the classical energy, the usual two-magnon terms, as well as three- and four-magnon interactions. We write
\begin{equation}
  H \approx N S(S+1) \epsilon_{\rm cl} + S H_2 + S^{1/2} H_3 + S^0 H_4,
\end{equation}
where the classical energy per site, $\epsilon_{\rm cl}$, is defined as
\begin{equation}
  \epsilon_{\rm cl}~\equiv~\frac{1}{2N_s}\sum_{\alpha\alpha'} \sum_{\vec{\delta}} \mathcal{J}^{00}_{\vec{\delta},\alpha\alpha'},
\end{equation}
\begin{widetext} 
and we have defined the individual pieces in symmetrized form as
\begin{subalign}
  {H}_2 &= \frac{1}{2}\sum_{\alpha\beta}\sum_{\vec{k}} \left[
    {A}_{\vec{k}}^{\alpha\beta} \h{a}_{\vec{k}\alpha}{a}^{}_{\vec{k}\beta} +
    {A}_{-\vec{k}}^{\beta\alpha} {a}^{}_{-\vec{k}\alpha}\h{a}_{-\vec{k}\beta} +
           \left( 
          {B}^{\alpha\beta}_{\vec{k}}\h{a}_{\vec{k}\alpha}\h{a}_{-\vec{k}\beta} +
          \cb{B}^{\alpha\beta}_{\vec{k}}{a}^{}_{-\vec{k}\beta}{a}^{}_{\vec{k}\alpha}\right)
          \right],\\
  {H}_3 &= \frac{1}{2!} \frac{1}{\sqrt{N_c}}\sum_{\alpha\beta\mu}\sum_{\vec{k}\vec{k}'} \left[
          {T}_{\vec{k}\vec{k}'}^{\alpha\beta\mu} \h{a}_{\vec{k}\alpha} \h{a}_{\vec{k}'\beta} {a}^{}_{\vec{k}+\vec{k}',\mu}+
          \cb{T}_{\vec{k}\vec{k}'}^{\alpha\beta\mu} \h{a}_{\vec{k}+\vec{k}',\mu}  {a}^{}_{\vec{k}'\beta}{a}^{}_{\vec{k}\alpha}
                \right],  \\
  {H}_4 &= \frac{1}{N_c}\sum_{\alpha\beta\mu\nu} \sum_{\vec{k}\vec{k}'\vec{q}}\left[
    \frac{1}{(2!)^2} {V}_{\vec{k}\vec{k}[\vec{q}]}^{\alpha\beta\mu\nu} \h{a}_{\vec{k}+\vec{q},\alpha}\h{a}_{\vec{k}'-\vec{q},\beta} {a}^{}_{\vec{k}'\mu} {a}^{}_{\vec{k}\nu} + 
                \frac{1}{3!} \left(
                {D}_{\vec{k}\vec{k}'\vec{q}}^{\alpha\beta\mu\nu}\h{a}_{\vec{k}\alpha}\h{a}_{\vec{k}'\beta} \h{a}_{\vec{q}\mu} {a}^{}_{\vec{k}+\vec{k}'+\vec{q},\nu} +\hc
                \right)
                \right]. 
\end{subalign}
where $N_s$ is the number of sublattices, $N$ is the total number of sites and we have defined the Fourier transforms of the bosons as $a_{\vec{k}\alpha} \equiv N_c^{-1/2} \sum_{\vec{r}} e^{-i\vec{k}\cdot\vec{r}} a_{\vec{r}\alpha}$ where $N = N_c N_s$. Similarly, we have defined the Fourier transforms of the local exchange matrices as 
$
\mathcal{J}^{\mu\mu'}_{\vec{k},\alpha\alpha'} \equiv \sum_{\vec{\delta}} \mathcal{J}^{\mu\mu'}_{\vec{\delta},\alpha\alpha'}e^{i\vec{k}\cdot\vec{\delta}} 
$.
In terms of these local exchange matrices [Eq.~(\ref{eq:lswt:exchange})] one can write              
\begin{subequations}
\label{eq:vertices}
\begin{align}
& {A}_{\vec{k}}^{\alpha\beta} =
    \mathcal{J}^{+-}_{\vec{k},\alpha\beta} - \delta_{\alpha\beta}\sum_\mu \mathcal{J}^{00}_{\vec{0},\alpha\mu} , \label{eq:vertices:A}\\  & {B}_{\vec{k}}^{\alpha\beta}
  =
    \mathcal{J}^{++}_{\vec{k},\alpha\beta},  \label{eq:vertices:B}\\ & {T}^{\alpha\beta\mu}_{\vec{k}\vec{k}'}
   = -\left[
    \delta_{\alpha\mu}\mathcal{J}^{+0}_{\vec{k}',\beta\alpha} + \delta_{\beta \mu} \mathcal{J}^{+0}_{\vec{k},\alpha\beta}   \right],\label{eq:vertices:T}\\
 & {V}^{\alpha\beta\mu\nu}_{\vec{k}\vec{k}'[\vec{q}]}
   =
    \left(\delta_{\alpha\mu}\delta_{\beta\nu}\mathcal{J}^{00}_{\vec{k}-\vec{k}'+\vec{q},\alpha\beta}+
    \delta_{\alpha\nu}\delta_{\beta\mu}\mathcal{J}^{00}_{\vec{q},\alpha\beta}\right)- 
    \left(
    \delta_{\mu\nu}\delta_{\mu\beta}\mathcal{J}^{+-}_{\vec{k}+\vec{q},\alpha\nu}+
    \delta_{\alpha\beta}\delta_{\alpha\mu}\mathcal{J}^{+-}_{\vec{k},\alpha\nu}
    \right),\label{eq:vertices:V} \\
 & {D}^{\alpha\beta\mu\nu}_{\vec{k}\vec{k}'\vec{q}}
  =
   -\frac{3}{4} \left(
    \delta_{\alpha\mu}\delta_{\alpha\nu} \mathcal{J}^{++}_{\vec{k}',\beta\alpha}+
    \delta_{\mu\beta}\delta_{\nu\beta} \mathcal{J}^{++}_{\vec{k},\alpha\beta}\right),\label{eq:vertices:D}
\end{align}
\end{subequations}
where the four-magnon vertices have been left unsymmetrized for brevity. 
As $1/S \rightarrow 0$, the interactions encoded in $H_3$ and $H_4$ can be included perturbatively as they are $O(S^{-1/2})$ and $O(S^{-1})$ with respect to the \ac{LSWT} parts.
\end{widetext}

\begin{figure}[bp]
  \centering
  \includegraphics[width=0.95\columnwidth]{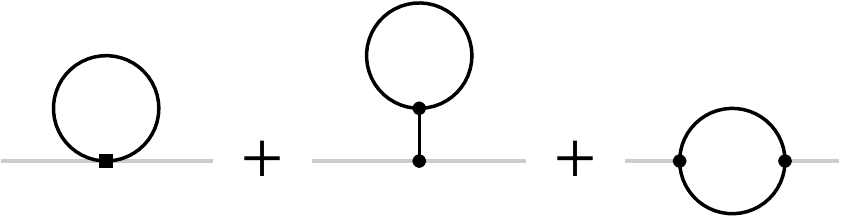}
  \caption{\label{fig:diagrams}
    The three classes of diagram contributing at leading order to the magnon self-energy, $\mat{\Sigma}^R(\vec{k},\omega)$. Propagators are free Green's functions (denoted as $\textbf{---}$) including both normal and anomalous parts. Each diagram is connected to two external legs (denoted as $\textcolor{cG}{\textbf{---}}$). Single four-magnon interaction vertices (denoted as {\scriptsize $\blacksquare$}) and pairs of three-magnon interaction vertices (denoted as $\bullet$) are included [see Eqs.~(\ref{eq:vertices:T}), (\ref{eq:vertices:V}) and (\ref{eq:vertices:D})]}
\end{figure}

Mainly, we are interested in the dynamical structure factor, defined as
\begin{equation}
  \mathcal{S}_{\nu\nu'}(\vec{q},\omega) \equiv \int d\omega\ e^{i\omega t} \avg{\mu^{\nu}_{-\vec{q}}(t) \mu^{\nu'}_{\vec{q}}},
\end{equation}
where the real-time, Fourier transformed magnetic moment operators are defined 
$
  \vec{\mu}_{\vec{q}}(t) \equiv \sum_r e^{i\vec{q}\cdot\vec{r}}e^{iHt}  \vec{\mu}_{r} e^{-iHt}
$.
A more useful form includes a polarization factor
\begin{equation}
\label{eq:dyn}
\mathcal{S}(\vec{q},\omega) \equiv \sum_{\nu\nu'} \left(\delta_{\nu\nu'} - \hat{q}_\nu \hat{q}_{\nu'}\right)\mathcal{S}_{\nu\nu'}(\vec{q},\omega),
\end{equation}
as this can be directly related to the intensity measured in an inelastic neutron scattering experiment~\cite{lovesey1984theory} via
\begin{equation}
  \label{eq:intensity}
  I(\vec{q},\omega) \propto F(\vec{q})^2 \mathcal{S}(\vec{q},\omega),
\end{equation}
where $F(\vec{q}) \equiv F(|\vec{q}|)$ is the atomic form factor of \rth{Yb}~\cite{inttablesc}.

The dynamical structure factor at $O(1/S^2)$ requires the computation of the magnon Green's function as well as several higher-order correlation functions.  It is useful to consider three distinct pieces: the transverse-transverse part, which involves only the magnon Green's function, $\mat{{G}}^R(\vec{k},\omega)$, the transverse-longitudinal parts, which involve three-magnon correlation functions, and the longitudinal-longitudinal parts which involve four-magnon correlation functions~\cite{mourigal2013dynamical}.  While the transverse-transverse part has $O(1/S)$ contributions, the other two parts appear first at $O(1/S^2)$. We note that the Green's function also appears in the transverse-longitudinal part of the dynamical structure factor, while the longitudinal-longitudinal part involves only the free magnon Green's function at leading order~\cite{mourigal2013dynamical}. Typically, the transverse-longitudinal and longitudinal-longitudinal parts are small relative to the leading transverse-transverse contributions.  The central ingredient is then the (retarded) magnon Green's function~\cite{ripka1986quantum}
\begin{align}
  \label{eq:nlswt:green}
  \mat{G}^R(\vec{k},\omega) & \equiv  \left[
  \mat{\sigma}_3(\omega+i0^+)-S\mat{M}_{\vec{k}}-\mat{\Sigma}^R(\vec{k},\omega)
  \right]^{-1}.
  \end{align}
where $\mat{\sigma}_3$ is a block Pauli matrix, $\mat{M}_{\vec{k}}$ is the (usual) \ac{LSWT} dispersion matrix
\begin{equation}
  \mat{M}_{\vec{k}} =\left(\begin{array}{cc}
                        \mat{A}_{\vec{k}} & \mat{B}_{\vec{k}} \\
                        \cc{\mat{B}}_{-\vec{k}} & \cc{\mat{A}}_{-\vec{k}}
  \end{array}\right),
\end{equation}
where $\mat{A}^{}_{\vec{k}} = \h{\mat{A}}_{\vec{k}}$ and $\vec{B}^{}_{\vec{k}} = \trp{\mat{B}}_{-\vec{k}}$ are defined in Eqs.~(\ref{eq:vertices:A}) and (\ref{eq:vertices:B}).

The self-energy, $\mat{\Sigma}^R(\vec{k},\omega)$, is generated by interactions, first at $O(S^0)$. As for the Green's function, this contains both normal and anomalous parts and sublattice indices.  The relevant diagrams at $O(S^0)$ in the magnon interactions are given in Fig.~\ref{fig:diagrams}.  We identify two distinct types of contributions to the self-energy: static (frequency independent) and dynamic (frequency dependent).  The static contributions arise from Hartree-Fock-like diagrams involving the four-magnon interactions as well as tadpole-like diagrams arising from the three-magnon interaction. The dynamic contributions arise purely from the three-magnon terms.  In addition to renormalizing the one-magnon spectrum they are also responsible for magnon decay~\cite{chubukov1994large,chernyshev2009triangular}, endowing the one-magnon states with finite lifetimes when kinematically allowed.

In practice, we first find the classical ground state numerically and compute the \ac{LSWT} energies and eigenvectors on a grid of wave-vectors corresponding to a finite system of $N=4L^3$ sites with periodic boundary conditions. For static quantities, typically we find $L=8$ is sufficient to reach convergence, given the (generically) gapped spectrum (Sec.~\ref{sec:competing-phases}). For dynamical quantities, we use $L=32$ when considering individual wave-vectors (Sec.~\ref{sec:cuts-001}, \ref{sec:cuts-1m10} and \ref{sec:cuts-111}), and $L=16$ or $L=24$ when integrating over some range in momentum space (Sec.~\ref{sec:direct-comparison} and \ref{sec:powder}).  

We then use these quantities to compute the required integrals in the self-energy diagrams numerically, working exclusively at zero temperature. These matrices then go into the Green's function [Eq.~(\ref{eq:nlswt:green})] along with a finite broadening $0^+ \rightarrow \delta \sim O(10^{-3})$. The matrix inversion can then be carried out numerically and the Green's function recovered; this effectively is a full solution of Dyson's equation and avoids some of the issues discussed in \citet{zhitomirsky2013colloquium} when trying to solve this equation perturbatively. This then yields the transverse-transverse part of the dynamical structure factor. Similar strategies are used for the transverse-longitudinal and longitudinal-longitudinal parts of the spin structure factor, with some (ad-hoc) resummations carried out in the $O(1/S^2)$ corrections to the transverse-transverse and transverse-longitudinal parts to avoid introducing additional spurious poles, as described (for example) in ~\citet{mourigal2013dynamical}.

Throughout we fix $S=1/2$, as appropriate for \yto{}, extrapolating from the asymptotic limit $S \rightarrow \infty$ where spin-wave theory is controlled.  Due to this choice our spin-wave expansion can break down, suggesting instabilities that appear only at small $S$, as we will see explicitly in Sec.~\ref{sec:competing-phases} and Sec.~\ref{sec:critical-fields}. This typically entails one of the (renormalized) excitation branches going soft, then continuing to negative (unphysical) energies. We also do not attempt to introduce self-consistency in the self-energy, be it in the static or dynamic parts; we will discuss some aspects of this in Sec.~\ref{sec:disc}. The results presented below, referred to as ``\ac{NLSWT}'', are computed following the above framework.


\section{Competing phases and renormalized phase boundaries}
\label{sec:competing-phases}
We begin with the effects of magnon interactions on the stability
of the \ac{SFM} state itself. Just as the softening of modes in \ac{LSWT} suggest a classical instability, corrections to these phase boundaries can be obtained from \ac{NLSWT} by considering where the (renormalized) spectrum goes soft. We explore these corrections in two contexts: in the location of the boundary between the SFM and $\Gamma_5$ phases in the anisotropic exchange model [Eq.~(\ref{eq:model})], and for the critical fields that separate the low-field \ac{SFM} and the high-field polarized phase for two high symmetry field directions.
\subsection{Splayed ferromagnet - $\Gamma_5$ phase boundary}
\label{sec:phase}
\begin{figure}[tp]
\centering
\includegraphics[width=\columnwidth]{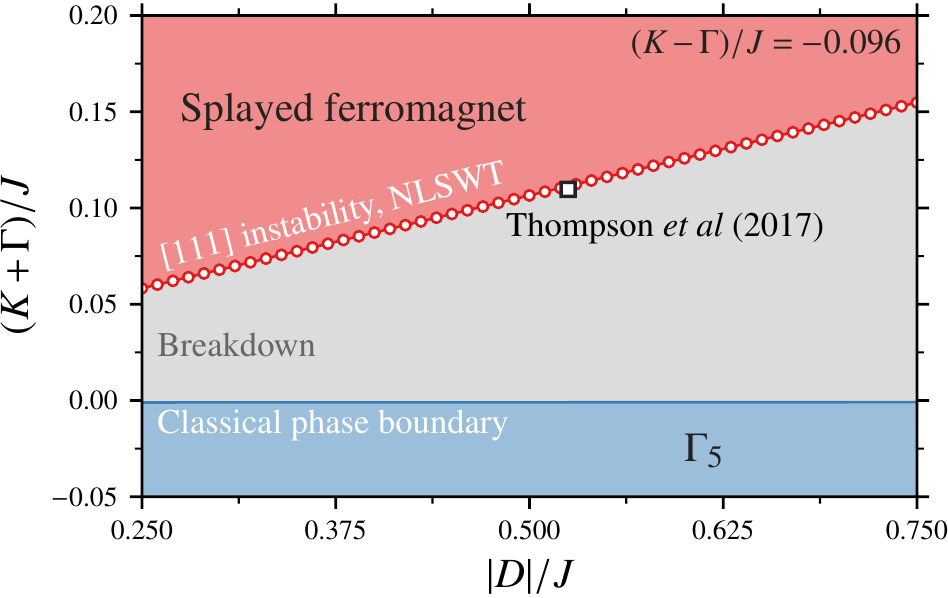}
\caption{
\label{fig:phase}
    Phase diagram in a region of nearest neighbor exchange parameters near the boundary of  the \acf{SFM} and $\Gamma_5$ phases, as a function of the strength  of (dual) Dzyaloshinskii-Moriya coupling,  $|D|/J$ and the (dual) symmetric anisotropic couplings $(K+\Gamma)/J$. The remaining parameter is chosen so that the exchanges of \citet{thompson2017} live in this plane (as shown), with  $(K-\Gamma)/J = -0.096$. Classically, the $\Gamma_5$ manifold is the ground state for $(K+\Gamma)/J \lesssim 0$, and the \ac{SFM} states for  $(K+\Gamma)/J \gtrsim 0$. Magnon interactions lead to an instability of the \ac{SFM} at larger $(K+\Gamma)/J$,  with spin-wave theory breaking down until the classical boundary at $(K+\Gamma)/J\sim 0$. For the parameters of Eq.~(\ref{eq:exchange-dual}), this is a shift of $+0.03 \meV$. The exchange parameters of \citet{thompson2017} lie very close to the renormalized, \acf{NLSWT} phase boundary.}
\end{figure}

We first consider the zero temperature phase diagram of the model [Eq.~(\ref{eq:model})] in the vicinity of the empirically determined exchange parameters of \parThompson{}. These are more usefully written in the dual global frame (see App.~\ref{app:conventions}), in terms of the (dual) Heisenberg ($J$), Kitaev ($K$), symmetric off-diagonal ($\Gamma$) and Dzyaloshinskii-Moriya ($D$) exchanges~\cite{PhysRevB.98.054408}.  In this language the parameters of \parThompson{} are
\begin{align}
\label{eq:exchange-dual}
J &= +0.272 \meV, & D &=-0.143 \meV, \nonumber\\ K &= +0.002 \meV, &  \Gamma &= +0.0279 \meV.
\end{align}
For $J>0$, $D<0$ and $|K|,|\Gamma| \ll J,|D|$, as is relevant here, the phase boundary between the \ac{SFM} and $\Gamma_5$ phase is $K+\Gamma \approx 0$~\cite{PhysRevB.98.054408,rau2019}. In Fig.~\ref{fig:phase} we show a  section of the $(K+\Gamma)/J$-$D/J$ plane containing the parameters of \parThompson{}. The classical phase diagram contains two phases - the \ac{SFM} corresponding to the ground state of \yto{} and the $\Gamma_5$ phase that hosts an accidentally degenerate $U(1)$  manifold of states. Quantum fluctuations lift the $\Gamma_5$ degeneracy~\cite{Jaubert2015,yan2017}  selecting the discrete set of $\psi_2$ or $\psi_3$ states (not shown)~\cite{poole2007magnetic}.  The empirically determined parameters for \yto{} lie close to this phase boundary as has been remarked upon previously~\cite{Jaubert2015,thompson2017,yan2017}. 

Carrying out \ac{NLSWT} about the classically stable \ac{SFM} state shows that the lowest spin wave mode goes soft before reaching the classical phase boundary. This line of instability is shown in Fig.~\ref{fig:phase}, reducing the range over which the \ac{SFM} state appears in the $(K+\Gamma)/J$ direction. In the region between the line of instability and the classical phase boundary the state is \emph{a priori} unknown. However, given the unstable mode appears with intensity at the $[111]$ wave-vector associated with the $\Gamma_5$ manifold, it is plausible that the phase that lies within this region is one of the order-by-quantum-disorder selected states $\psi_2$ or $\psi_3$. We stress that the line of instability potentially \emph{over}-estimates the range of the \ac{SFM} state, with a first order transition possible already at larger $(K+\Gamma)/J$. In addition, the effects of quantum fluctuations may introduce a finite gap to the soft mode, even at the phase boundary~\cite{rau2018pseudo}.

We note that the parameters of \parThompson{} lie very close to the line of instability, just on the unstable (nominally $\Gamma_5$) side. This proximity reaffirms some of the arguments of \citet{Jaubert2015}, where it was argued that the parameters of \citet{ross2011} were pushed closer to the \ac{SFM}-$\Gamma_5$ phase boundary by quantum fluctuations. The calculations presented here suggest that this remains true for the parameters of \citet{thompson2017}; the proximity to this phase boundary likely plays an important role in understanding the physics of \yto{}.

\subsection{Critical magnetic fields}
\label{sec:critical-fields}

\begin{figure} 
        \includegraphics[width=0.95\columnwidth]{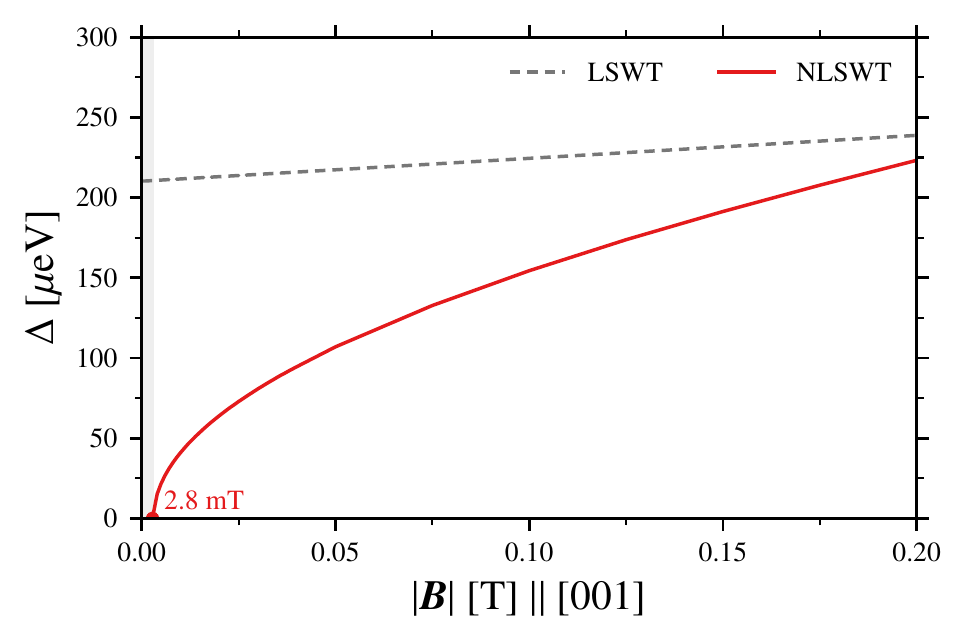}
        \caption{\label{fig:stab-field}
          Spin-wave gap ($\Delta$) at wave-vector $[000]$ computed in \acf{LSWT} and \acf{NLSWT} as a function of applied $[001]$ magnetic field for the parameters of \parThompson{}. The \acl{SFM} becomes unstable at a small (but finite) field $\sim 2.8\mT$, due to its close proximity to the phase boundary to the $\Gamma_5$ phase (see Fig.~\ref{fig:phase}). The sharp softening of this mode only occurs in a small field range below $|\vec{B}| \lesssim 0.2\T$.  The region where spin-wave theory has broken down is indicated.}
\end{figure}

We now consider the behavior of the phase boundaries as a function of magnetic field. First, let us note that the instability of the parameters of \parThompson{} that appears at zero-field can be resolved by applying a very small $[001]$ magnetic field of the order of $\sim 10^{-3}\T$. As shown in Fig.~\ref{fig:stab-field}, this opens a small gap for $|\vec{B}| \gtrsim 3 \mT$ with intensity at the $[111]$ wave-vector. This can be contrasted with \ac{LSWT}, where there is a finite $\sim 0.2 \meV$ gap down to zero field. We thus see that, at small $[001]$ fields, there is a large renormalization of the $[111]$ mode, pushing it to low energy. The strong renormalization of this mode has mostly disappeared by $|\vec{B}| \sim 0.2\T$, as shown in Fig.~\ref{fig:stab-field}. The smallness of the required stabilization field reflects the extreme proximity of these parameters to the line of instability in the phase diagram discussed in Sec.~\ref{sec:phase}, and shown in Fig.~\ref{fig:phase}. We note that this zero-field instability is \emph{not} present in the parameters of \citet{ross2011}, due to its distance from the boundary, but \emph{is} present for the parameters of \citet{Robert2015}, which are (in some sense) closer to the classical boundary than those of \citet{thompson2017} (and thus are in the region where \ac{NLSWT} breaks down).

\begin{figure}
  \begin{overpic}[width=0.95\columnwidth]{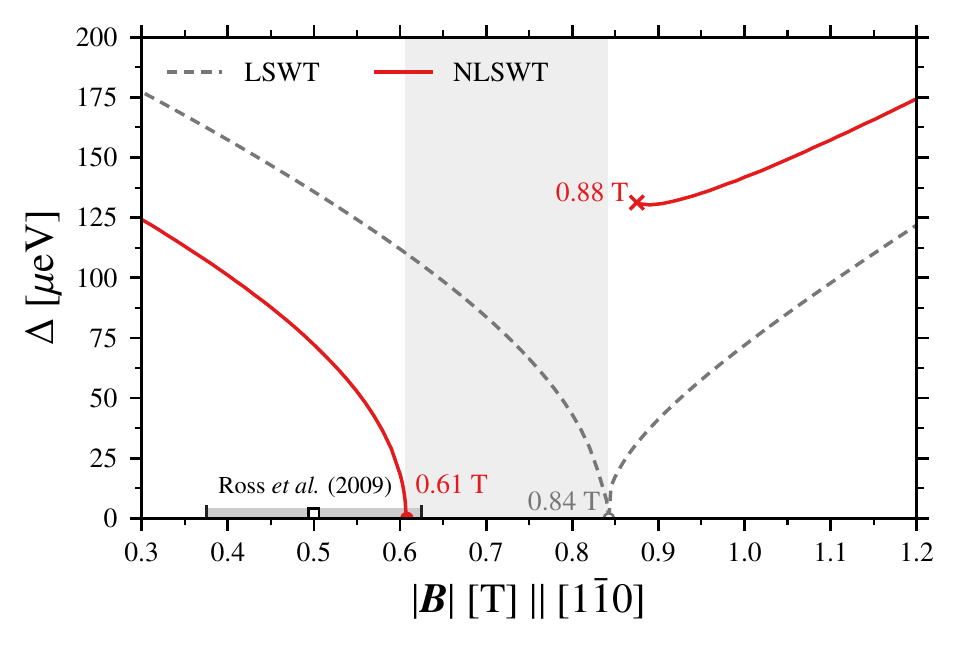}        
    \put(90,15.5){(a)}
  \end{overpic}
  \begin{overpic}[width=0.95\columnwidth]{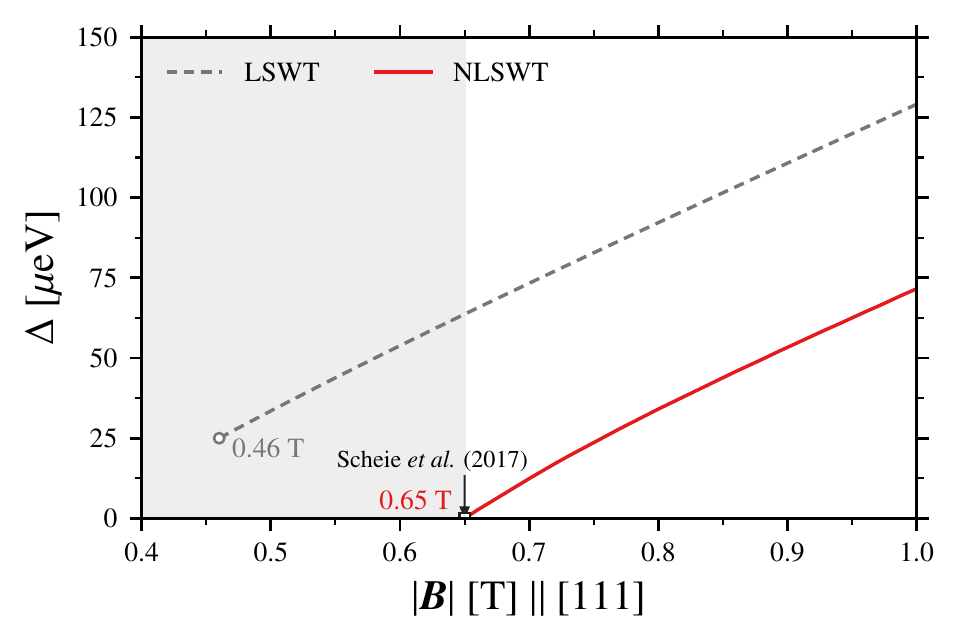}        
    \put(90,15.5){(b)}
  \end{overpic}
  \caption{\label{fig:crit-field}
    Spin-wave gap ($\Delta$) at wave-vector $[000]$ computed in \acf{LSWT} and \acf{NLSWT} as a function of applied (a) $[1\bar{1}0]$ and (b) $[111]$ magnetic fields for the parameters of \parThompson{}.  For $[1\bar{1}0]$ fields in \ac{NLSWT} there is an instability at $|\vec{B}| \sim 0.61\T$ out of the \acl{SFM} phase, not far from the experimental value, $|\vec{B}| \sim 0.5 \pm 0.125\T$, of \citet{Ross2009}.  For $[111]$ fields, \ac{NLSWT} becomes unstable at $|\vec{B}| \sim 0.65\T$, close to the experimentally determined field \citet{scheie2017} and above the classical critical field of $0.46 \T$.  The region where spin-wave theory has broken down is indicated.}
\end{figure}

In contrast to the $[001]$ field direction, there are field-induced phase transitions out of the \ac{SFM} state for both $[1\bar{1}0]$ and $[111]$ field directions, due to their remnant two-fold symmetry. To analyze what effects magnon interactions have on these phase boundaries, we track the evolution of the lowest lying mode at the $[000]$ wave-vector as a function of field. We restrict ourselves to the parameters of \parThompson{} throughout.

The result for $[1\bar{1}0]$ field is shown in Fig.~\subref{fig:crit-field}{(a)}. We see that coming down from the high-field phase the \ac{LSWT} gap closes at $|\vec{B}| \sim 0.84\T$, well above the $|\vec{B}| \sim 0.5 \pm 0.125 \T$~\footnote{The quoted error bars are our estimate due to the experiments being carried out for only a few fields near the critical field, namely at $|\vec{B}| = 0.25\T$, $0.5\T$ and $0.75\T$~\cite{Ross2009}.} determined experimentally~\cite{Ross2009} . The \ac{NLSWT} result breaks down artificially above the classical field, due to the softening the bare \ac{LSWT} spectrum. However from Fig.~\subref{fig:crit-field}{(a)} it is clear that the gap at $[000]$ is strongly enhanced, suggesting that the critical field is lowered by the effects of magnon interactions, as is found experimentally. Given that \ac{NLSWT} is limited by the classical phase boundaries, it is not directly possible to obtain a lowering of the critical field in this fashion.  To resolve this, we have also looked at this gap coming out of the ordered phase. We find that in \ac{NLSWT} the \ac{SFM} state has an instability at $[000]$ which goes soft at $|\vec{B}| \sim 0.61\T$ [see Fig.~\subref{fig:crit-field}{(a)}], somewhat above the experimental value but not quantitatively inconsistent.

The result for $[111]$ field is shown in Fig.~\subref{fig:crit-field}{(b)}. The classical transition field in this case, $|\vec{B}| \sim 0.46\T$, is lower than the $|\vec{B}| \sim 0.65\T$ that has been recently observed experimentally~\cite{scheie2017}. Both theoretically and experimentally this is expected to be a first-order transition, with the gap remaining open across the transition.  In contrast to the $[1\bar{1}0]$ case it should thus be possible to obtain an increase in the transition field from \ac{NLSWT} in the polarized phase. One finds that the gap at $[000]$ is strongly reduced by interaction effects, finally going soft very close to the experimental value of $|\vec{B}| \sim 0.65\T$. Given the first order nature of the transition seen experimentally, we expect the precise matching of these results to be somewhat accidental, with the true (theoretical) transition being first order at a field above the \ac{NLSWT} instability.

\section{Dynamical Structure Factor}
\label{sec:DSF} 

We now explore more broadly the interaction corrections from \ac{NLSWT} to the dynamical structure factor. We consider both the parameter sets of \citet{ross2011} and \citet{thompson2017} as well as fields in the $[001]$ (Sec.~\ref{sec:cuts-001}), $[1\bar{1}0]$ (Sec.~\ref{sec:cuts-1m10}) and $[111]$ (Sec.~\ref{sec:cuts-111}) directions. To obtain a global picture of the spectrum, we compute the dynamical structure factor along paths in momentum space between high symmetry points, that (mostly) lie in the plane perpendicular to the applied magnetic field. The results are shown in Fig.~\ref{fig:spaghetti-001} ($[001]$ field), Fig.~\ref{fig:spaghetti-1m10} ($[1\bar{1}0]$ field) and Fig.~\ref{fig:spaghetti-111} ($[111]$ field). For each we show the full dynamical structure factor, including the transverse-transverse, transverse-longitudinal and longitudinal-longitudinal components (as discussed in Sec.~\ref{sec:swt}). This result is convolved with a narrow Gaussian to improve visibility of  any sharp modes. In addition to the structure factor in \ac{NLSWT}, we also include the bare \ac{LSWT} one-magnon spectrum and the edges of the bare two-magnon continuum. 

After this overview, we consider two direct comparisons to experimental data, namely the single crystal data of \citet{thompson2017} (Sec.~\ref{sec:direct-comparison}) and the powder-averaged data of \citet{VPA2017} (Sec.~\ref{sec:powder}). For each case the integration over momentum space performed experimentally is reproduced theoretically with the form factor included [Eq.~(\ref{eq:intensity})].

\subsection{[001] magnetic field}
\label{sec:cuts-001}
\begin{figure*}
  \begin{overpic}[width=0.95\textwidth]{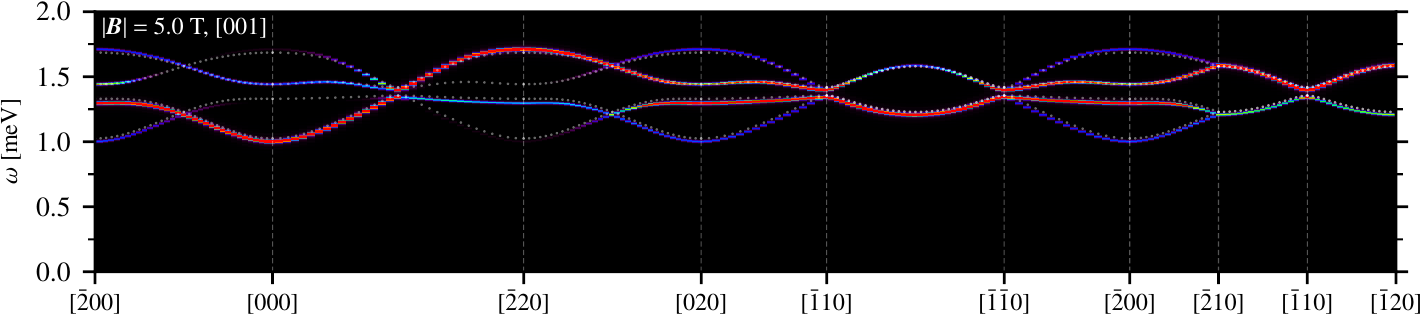}
    \put(95.5,19.5){\textcolor{white}{(a)}}
    \put(21,19.66){\textcolor{white}{\citetalt{thompson2017}}}
  \end{overpic}  
  \begin{overpic}[width=0.95\textwidth]{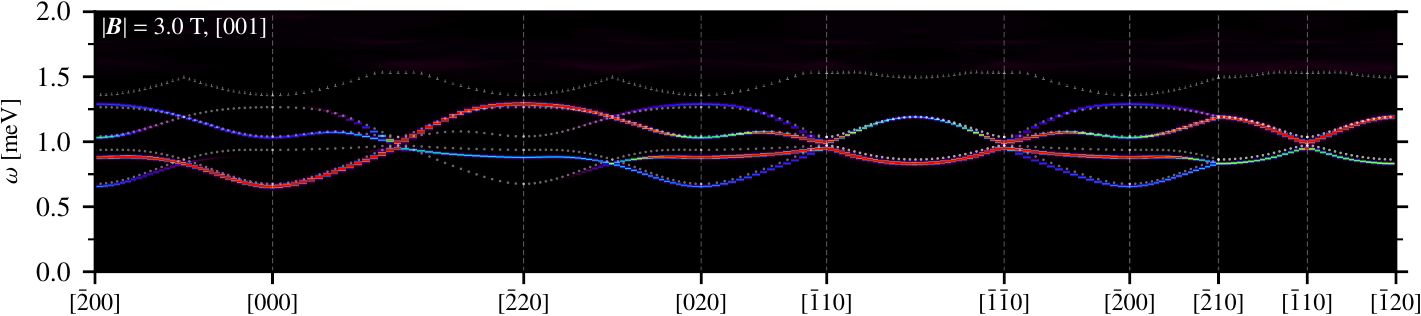}
    \put(95.5,19.5){\textcolor{white}{(b)}}
  \end{overpic}  
  \begin{overpic}[width=0.95\textwidth]{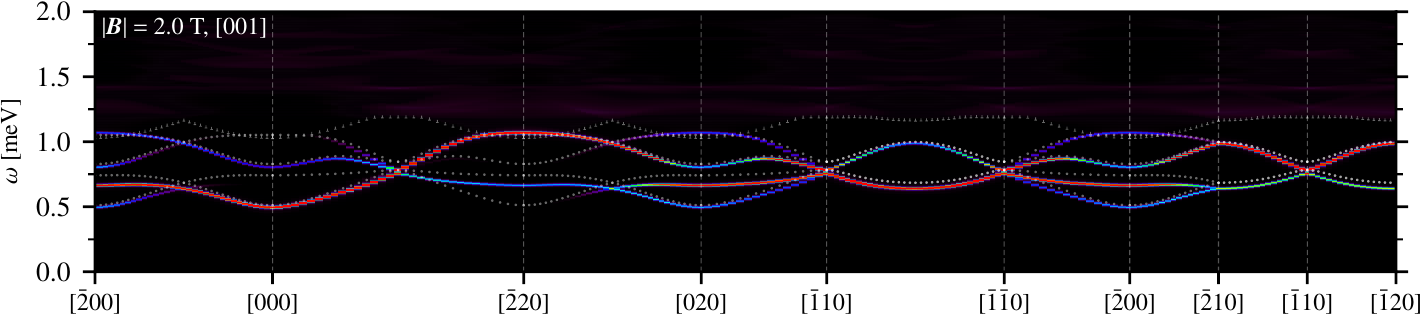}
    \put(95.5,19.5){\textcolor{white}{(c)}}
  \end{overpic}  
  \begin{overpic}[width=0.95\textwidth]{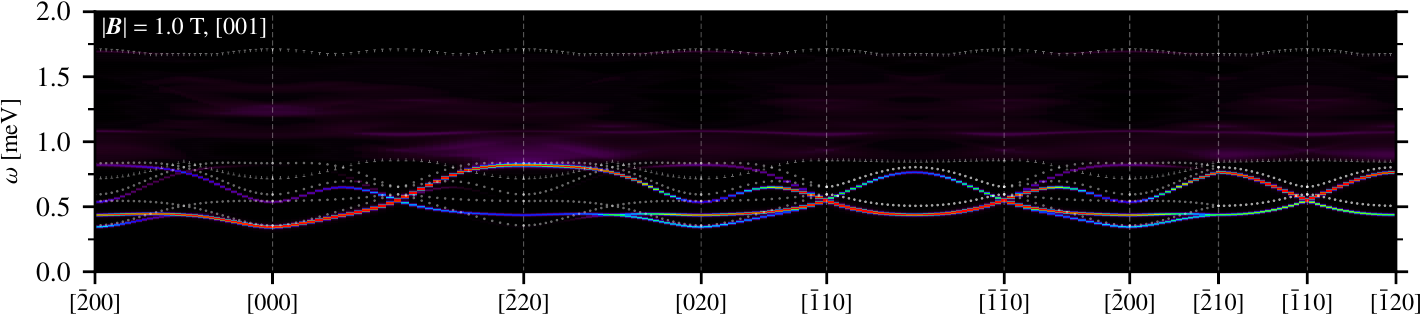}
    \put(95.5,19.5){\textcolor{white}{(d)}}
  \end{overpic}  
  \begin{overpic}[width=0.95\textwidth]{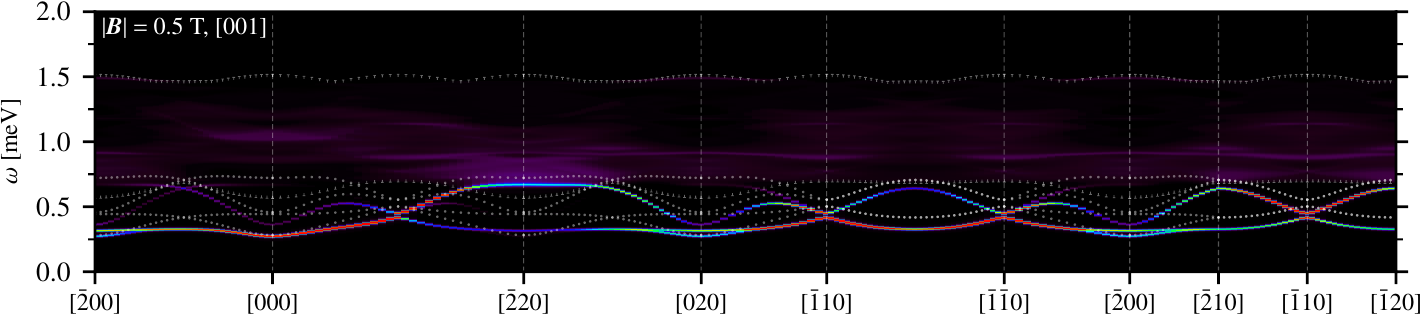}
    \put(95.5,19.5){\textcolor{white}{(e)}}
  \end{overpic}  
  \begin{overpic}[width=0.95\textwidth]{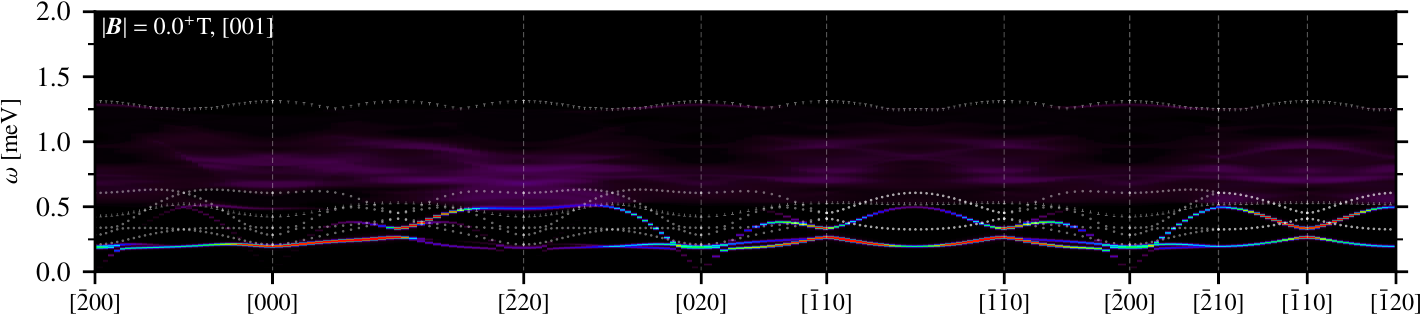}
    \put(95.5,19.5){\textcolor{white}{(f)}}
  \end{overpic}  
  \caption{\label{fig:spaghetti-001}
    (a-f) Dynamical structure factor in \acl{NLSWT} along a path in momentum space for a $[001]$ magnetic field of strength (a) $|\vec{B}|=5\T$, (b) $3\T$, (c) $2\T$, (d) $1\T$, (e) $0.5\T$ and (f) $0.0^+\T$ for the parameters of \citet{thompson2017}. Intensity shows $\mathcal{S}(\vec{q},\omega)$ [Eq.(\ref{eq:dyn})], while the dots mark the one-magnon energies in \acl{LSWT} and the upward pointing (downward pointing) symbols indicate the lower (upper) boundary  of the two-magnon continuum. Overall intensity is arbitrary, but consistent across panels. The instability at zero-field can be seen in the soft mode in (f), near $[020]$.}  
\end{figure*}
     
First, let us revisit the dependence of the excitations on a $[001]$ field, as was used by \citet{thompson2017} to determine the exchanges and $g$-factors. A sequence of fields, from $|\vec{B}| = 5\T$ down to $|\vec{B}|=0^+\T$ (to select a single domain) is shown in Fig.~\ref{fig:spaghetti-001} using the parameters of \citet{thompson2017}. Examining the \ac{LSWT} and \ac{NLSWT} results at $|\vec{B}| = 5\T$ (see Fig.~\subref{fig:spaghetti-001}{(a)}), we see that the two-magnon continuum is well-separated from the one-magnon states, and corrections due to magnon interactions are small. We conclude that \ac{LSWT} is sufficient to describe the spectrum for $|\vec{B}| \gtrsim 5\T$, and thus the fitting procedure used in \citet{thompson2017} is sensible, up to the level of precision implied in Fig.~\subref{fig:spaghetti-001}{(a)}.

As we lower the field, interaction effects become important, with significant differences between the \ac{LSWT} and \ac{NLSWT} spectrum becoming apparent for $|\vec{B}| \lesssim 3\T$.  These are the fields at which the two-magnon continuum begins to overlap with the one-magnon bands [see Fig.~\subref{fig:spaghetti-001}{(c)}]. Roughly speaking, one may rationalize the renormalizations as a kind of level repulsion between the one-magnon spectrum and the encroaching two-magnon continuum. When the one- and two-magnon states overlap, spontaneous magnon decay is kinematically allowed. Such decay is most apparent for fields $|\vec{B}|\lesssim 2\T$ near the upper bands at $[2\bar{2}0]$, as shown in Figs.~\subref{fig:spaghetti-001}{(d-f)}.  In such regions, the one-magnon excitations acquire a true linewidth, as opposed to the artificial width given to the (sharp) magnon modes outside these regions. As we lower the field further, this mode becomes more and more damped, as the two-magnon density of states increases at its energy.

Below $|\vec{B}| \lesssim 1\T$ the spectrum begins to stabilize with only small, quantitative changes appearing, as shown in Fig.~\subref{fig:spaghetti-001}{(e,f)}. In contrast to experiments, this calculation exhibits sharp magnon modes across most of momentum space, within the energy range $0.2\meV \lesssim \omega \lesssim 0.7\meV$, save for a few regions where decay is kinematically allowed.  We further note that the intense magnon mode at $0.2\meV$ near $[000]$ remains present (at roughly the same energy as in \ac{LSWT}), but is significantly flattened. Indeed, the \ac{NLSWT} corrections remove nearly all of the dispersion along the $[h00]$ direction, as can be seen in Fig.~\subref{fig:spaghetti-001}{(e)}. Similar magnon band flattening has been observed in interaction corrections to the spin one-half triangular lattice antiferromagnet \cite{zheng2006,starykh2006}. In addition to these sharp one-magnon modes, the intensity of the two-magnon continuum is also enhanced at low fields, appearing as broad intensity in the range $0.5 \meV \lesssim \omega \lesssim 1.0\meV$.

Finally, as discussed in Sec.~\ref{sec:critical-fields}, at strictly zero-field the \ac{SFM} state is unstable, with a mode going slightly soft at $[111]$. Remnants of this mode can be seen near $[\bar{2}00]$ and $[020]$ in Fig.~\subref{fig:spaghetti-001}{(f)}. By application of a small field ($|\vec{B}|\sim 3\mT$) this mode can be lifted to finite energy. The qualitative features of the spectrum discussed above are not strongly affected by the presence or absence of this mode, given its low density of states. However, it would appear as an intense, nearly gapped mode near $[111]$, signalling the proximity to the SFM-$\Gamma_5$ phase boundary (see Sec.~\ref{sec:disc} for further discussion).

\subsection{$[1\bar{1}0]$ magnetic field}
\label{sec:cuts-1m10}
\begin{figure*}
  \begin{overpic}[width=0.95\textwidth]{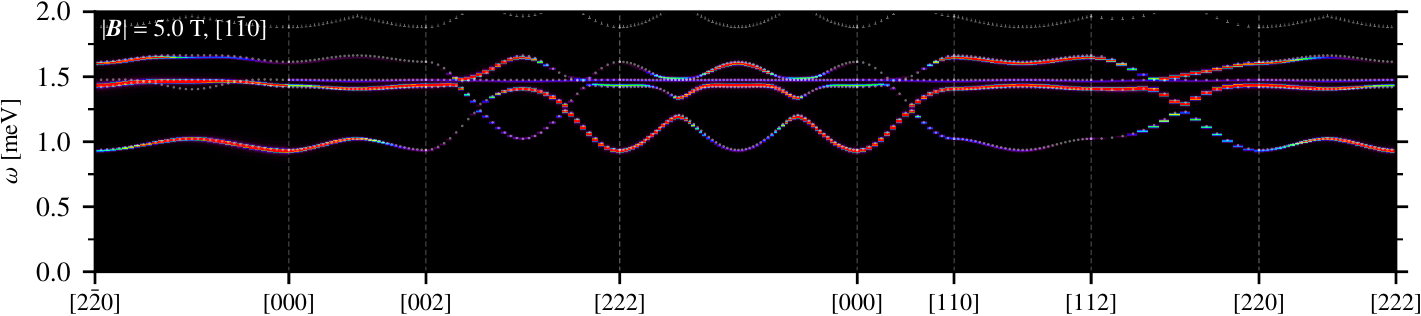} 
    \put(95.5,19.5){\textcolor{white}{(a)}}
    \put(21,19.66){\textcolor{white}{\citetalt{thompson2017}}}
  \end{overpic}
  \begin{overpic}[width=0.95\textwidth]{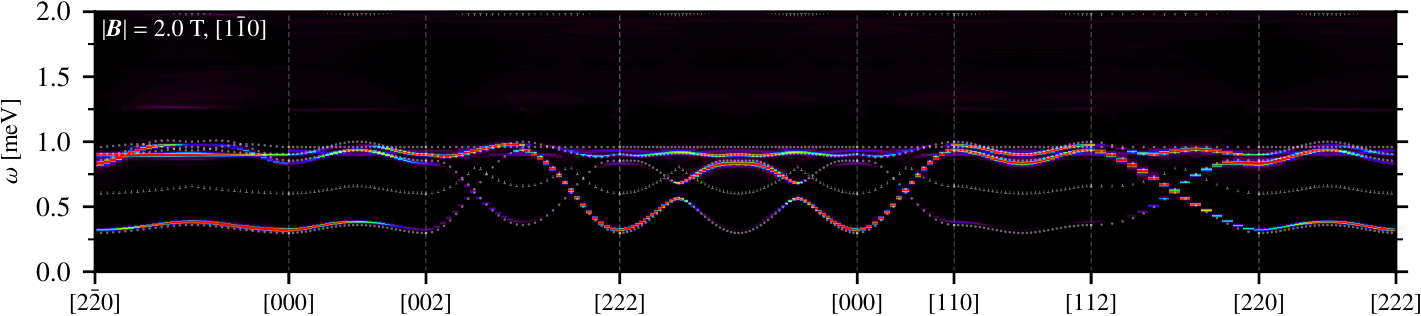} 
    \put(95.5,19.5){\textcolor{white}{(b)}}
  \end{overpic}
  \caption{\label{fig:spaghetti-1m10}
    (a-d) Dynamical structure factor in \acl{NLSWT} along a path in momentum space for a $[1\bar{1}0]$ magnetic field of strength (a) $|\vec{B}|=5\T$ and (b) $2\T$, for the parameters of \citet{thompson2017}. Intensity shows $\mathcal{S}(\vec{q},\omega)$ [Eq.(\ref{eq:dyn})], while the dots mark the one-magnon energies in \acl{LSWT} and the upward pointing (downward pointing) symbols indicate the lower (upper) boundary of the two-magnon continuum. Overall intensity is arbitrary, but consistent across panels.}
\end{figure*}

Next we consider the dynamical structure factor for an applied $[1\bar{1}0]$ field, as shown in Fig.~\ref{fig:spaghetti-1m10}. We focus on the two experimentally studied fields $|\vec{B}| = 2\T$ and $|\vec{B}| = 5\T$, as first studied in Ref.~\cite{ross2011}, both above the critical field in the polarized phase, using the parameters of \citet{thompson2017}.

As in the case of $[001]$, at $|\vec{B}|=5\T$ the \ac{NLSWT} calculation matches \ac{LSWT} almost perfectly, further justifying the use of \ac{LSWT} in fitting the model at these fields [Fig.~\subref{fig:spaghetti-1m10}{(a)}]. The applicability of \ac{LSWT} at the lower field, $|\vec{B}|=2\T$, is less clear on na\"ive grounds. From simple kinematic considerations one may worry about interaction effects becoming important, as the two-magnon continuum has begun to overlap with the one-magnon states, as shown in Figs.~\subref{fig:spaghetti-1m10}{(b)}. However, while there is some decay and some renormalization, its effects are rather small relative to (say) typical experimental resolution, and so for $[1\bar{1}0]$ even \ac{LSWT} is likely sensible at $|\vec{B}| \gtrsim 2\T$. We have confirmed that this is somewhat generic, also holding for the parameters of \citet{ross2011} (not shown).

\subsection{$[111]$ magnetic field}
\label{sec:cuts-111}
\begin{figure*}
  \begin{overpic}[width=0.95\textwidth]{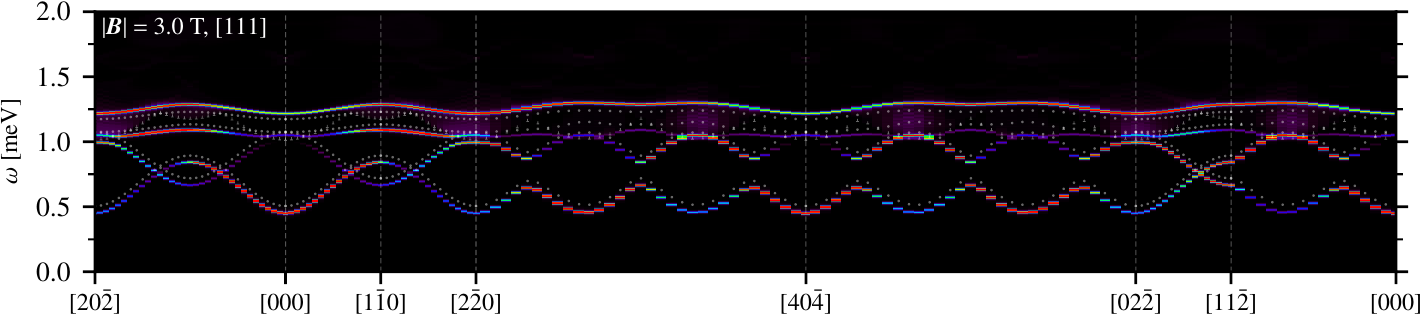}
    \put(95.5,19.5){\textcolor{white}{(a)}}
    \put(21,19.66){\textcolor{white}{\citetalt{thompson2017}}}
  \end{overpic}   
  \begin{overpic}[width=0.95\textwidth]{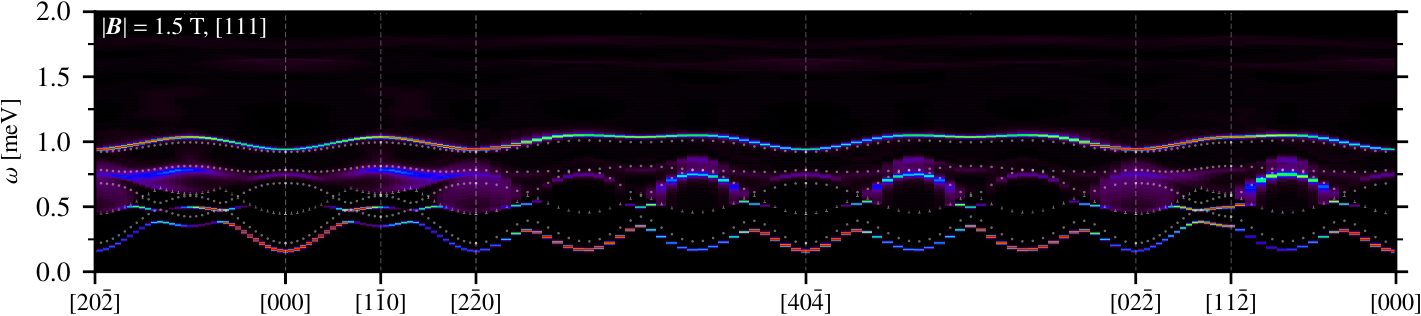}
    \put(95.5,19.5){\textcolor{white}{(b)}}
  \end{overpic}   
  \caption{\label{fig:spaghetti-111}
    (a,b) Dynamical structure factor in \acl{NLSWT} along a path in momentum space for $[111]$ magnetic fields of strength (a) $|\vec{B}|=3\T$ and (b) $1.5\T$ for the parameters of \citet{thompson2017}. Intensity shows $\mathcal{S}(\vec{q},\omega)$ [Eq.(\ref{eq:dyn})], while the dots mark the one-magnon energies in \acl{LSWT} and the upward pointing (downward pointing) symbols indicate the lower (upper) boundary  of the two-magnon continuum. Overall intensity is arbitrary, but consistent across panels.}
\end{figure*}

We now turn to results for the third high symmetry field direction $[111]$, which has been of interest recently~\cite{scheie2017}. Here we find considerable departures from \ac{LSWT} even at $|\vec{B}| = 3 \T$ [see Fig.~\subref{fig:spaghetti-111}{(a)}], well above the transition field of $|\vec{B}| \sim 0.65\T$~\cite{scheie2017}.  Even at this high field several of the one-magnon bands strongly intersect with the two-magnon continuum. The highest energy mode, originating from the pinned moment of the single sublattice distinguished by the field direction, lies entirely within the two-magnon continuum. This induces renormalization of the energies as well as spontaneous magnon decay, for example near [$2\bar{2}0]$ as shown in Fig.~\subref{fig:spaghetti-111}{(a)}.

Going down to $|\vec{B}| = 1.5\T$, still significantly above the transition field of $|\vec{B}| \sim 0.65\T$~\cite{scheie2017}, there are drastic corrections to the \ac{LSWT} spectrum. While lowest mode remains sharp, it receives large corrections, especially where it meets the second and third modes of the spectrum. These two modes are nearly completely destroyed by spontaneous magnon decay; over most of the wave-vectors shown in Fig.~\subref{fig:spaghetti-111}{(b)} there is either significant decay or significant renormalization, or both. The $[111]$ polarized phase thus appears to be an excellent setting to experimentally observe and study spontaneous magnon decay. We note that the highest energy band is relatively sharp and only weakly renormalized by magnon interactions. Finally, we have also confirmed that the features discussed above are also present in the parameters of \citet{ross2011} (not shown), and thus appear to be relatively generic for $[111]$ fields. However, given the small one-magnon gap at $1.5\T$, and thus the low-lying two-magnon continuum, details of the renormalization and induced decay do depend somewhat on the precise exchanges considered.

\subsection{Direct comparison to experiment in [001] field}
\label{sec:direct-comparison}

\begin{figure*}
  \centering
  \begin{overpic}[width=0.85\textwidth]{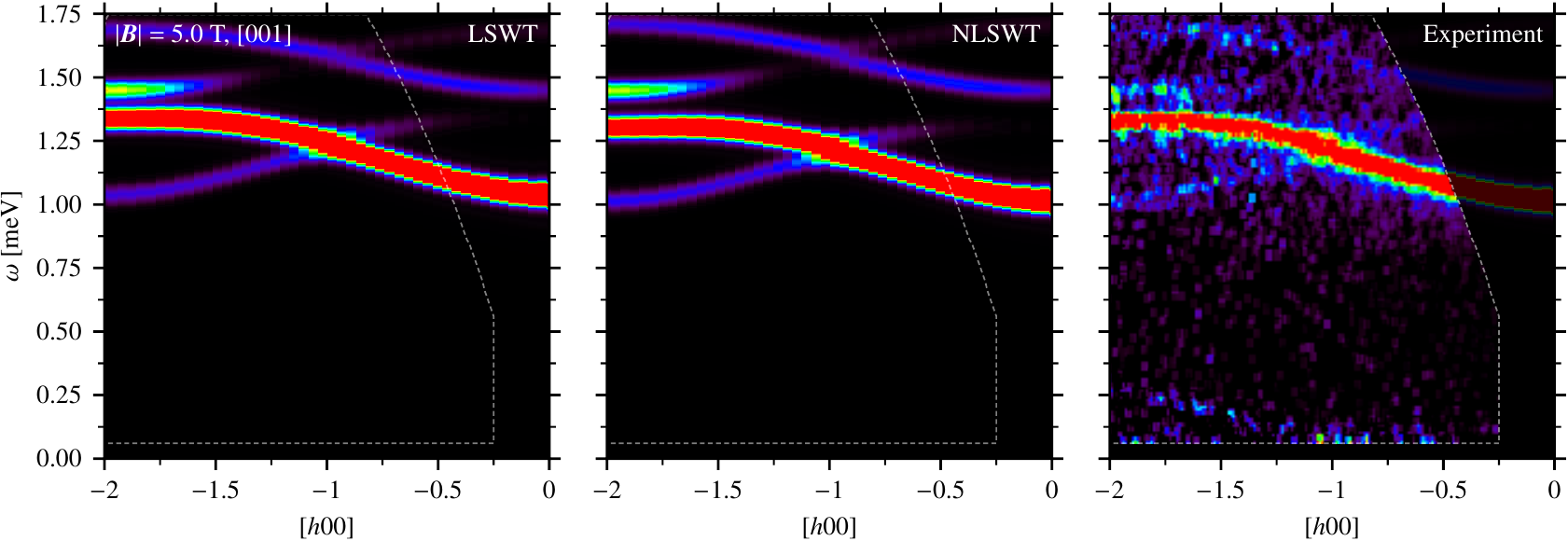}
    \put(-1,32.5){(a)}
  \end{overpic}\vspace{0.2cm}
  \begin{overpic}[width=0.85\textwidth]{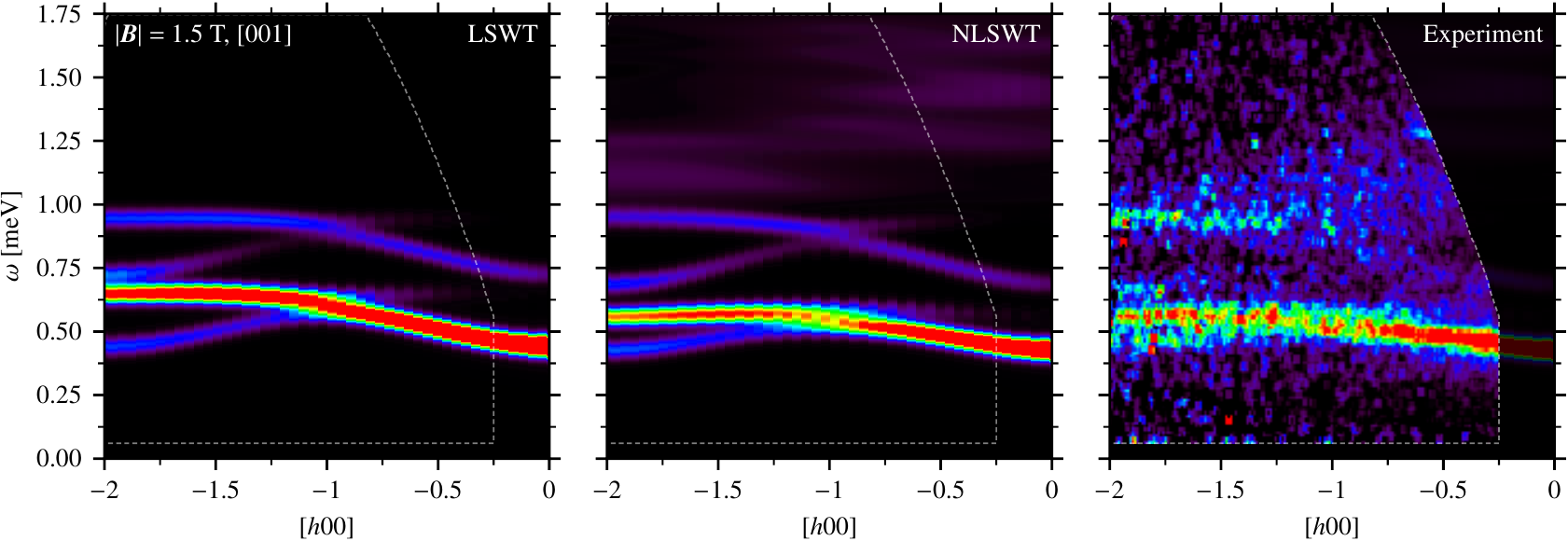}
    \put(-1,32.5){(b)}
  \end{overpic}\vspace{0.2cm}
  \begin{overpic}[width=0.85\textwidth]{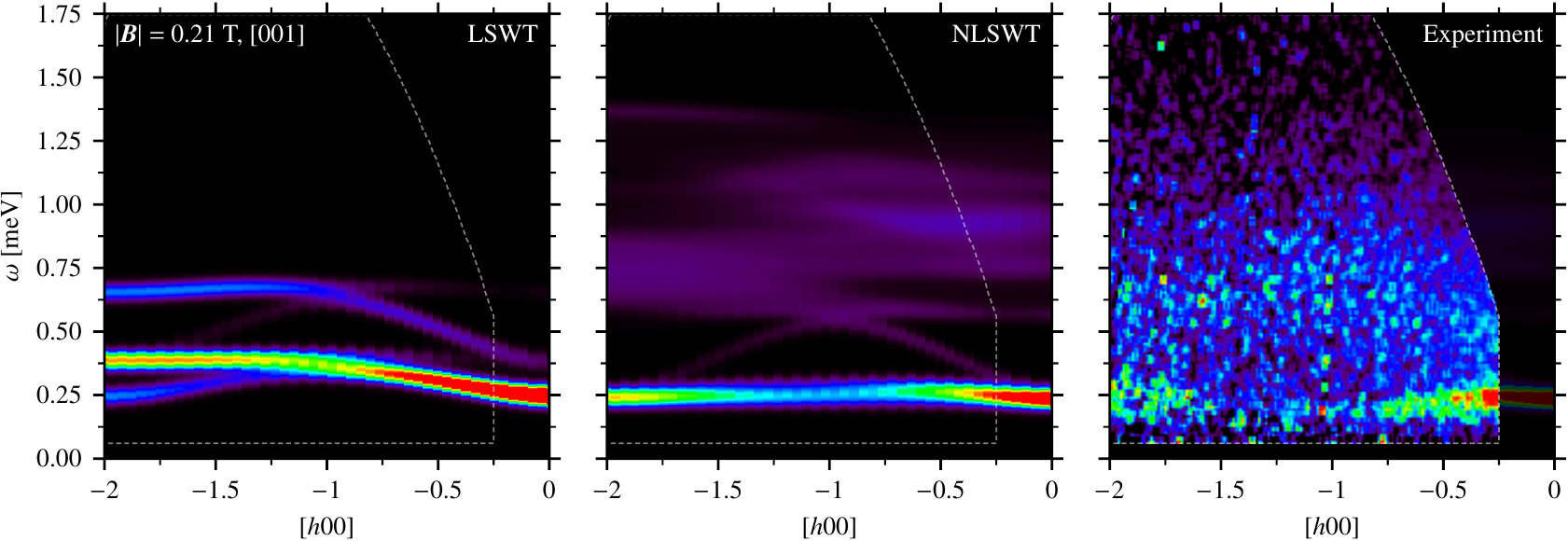}
    \put(-1,32.5){(c)}
  \end{overpic}\vspace{0.2cm}
  \begin{overpic}[width=0.85\textwidth]{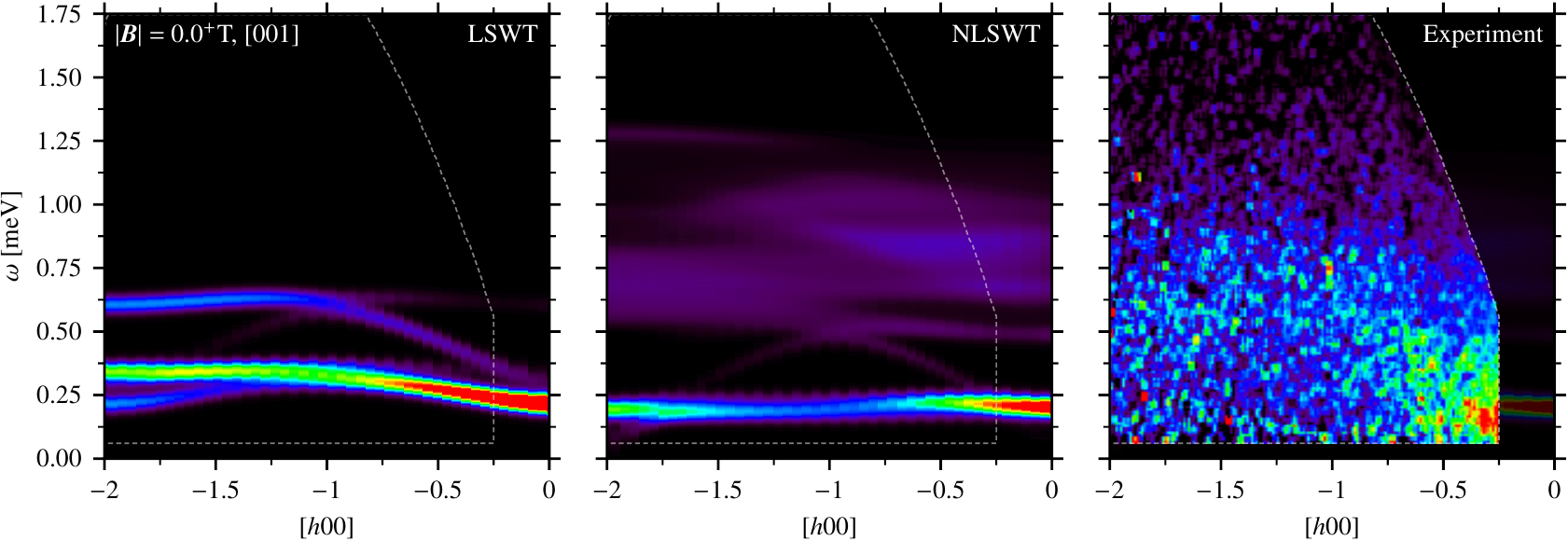}
    \put(-1,32.5){(d)}
  \end{overpic}
  \caption{\label{fig:exp100}
    (a-d) Comparison of inelastic neutron scattering intensity, $I(\vec{q},\omega)$ [Eq.~(\ref{eq:intensity})], in \acf{LSWT},  \acf{NLSWT} and in the experimental data of \citet{thompson2017} along $[h00]$ for four different fields along $[001]$, (a) $|\vec{B}| = 5\T$, (b) $1.5\T$, (c) $0.21\T$ and (d) $0.0^+\T$. Intensities in all plots are averaged  over wave-vectors along transverse directions over the range $l, k = [-0.2,0.2]$. Overall scale is chosen to match experimental data, and is consistent across Figs.~\ref{fig:exp100}, \ref{fig:exp1m10}, and \ref{fig:exp110}. The dashed lines indicate the boundaries of the experimental data, while features outside these boundaries in the experimental panels correspond to the \ac{NLSWT} calculation.}
 \end{figure*}
 
\begin{figure*}
  \centering
  \begin{overpic}[width=0.85\textwidth]{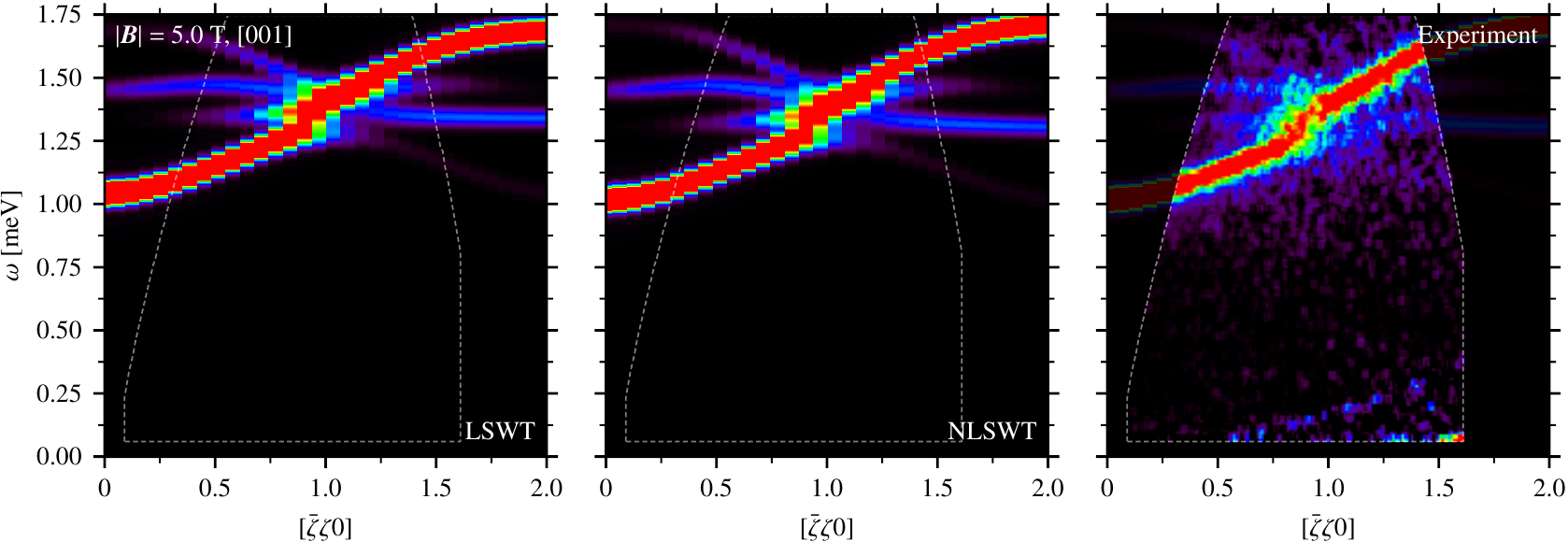}
    \put(-1,32.75){(a)}
  \end{overpic}\vspace{0.2cm}
  \begin{overpic}[width=0.85\textwidth]{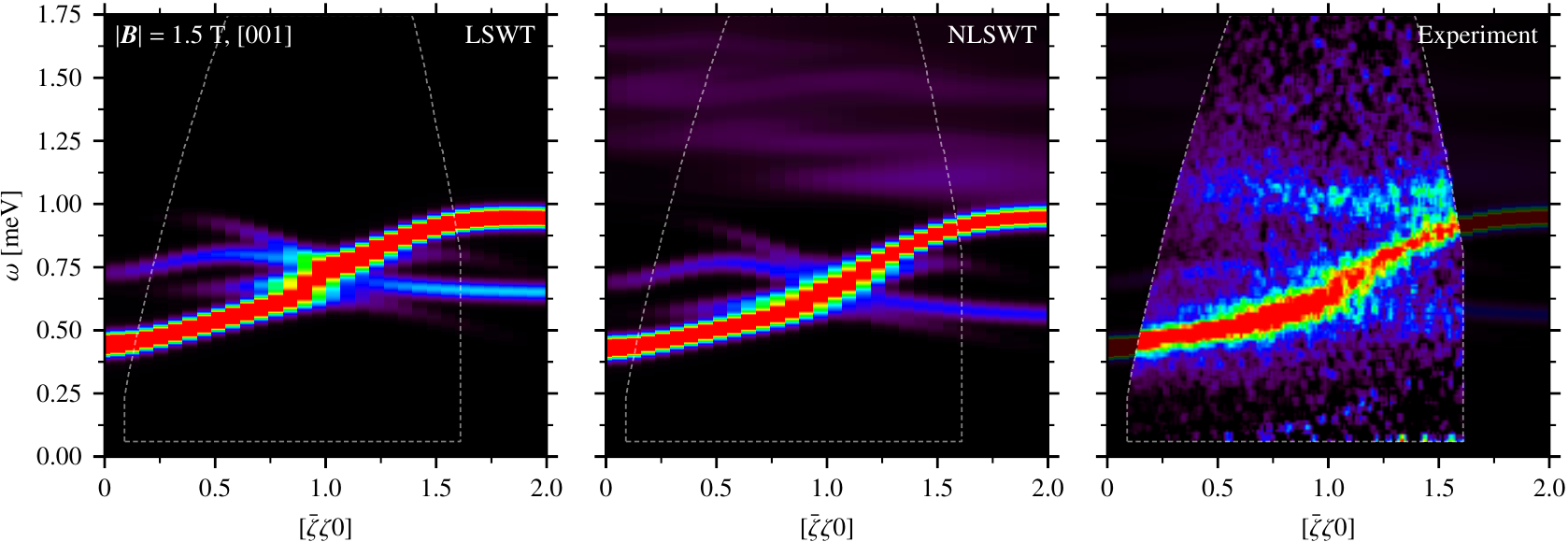}
    \put(-1,32.75){(b)}
  \end{overpic}\vspace{0.2cm}
  \begin{overpic}[width=0.85\textwidth]{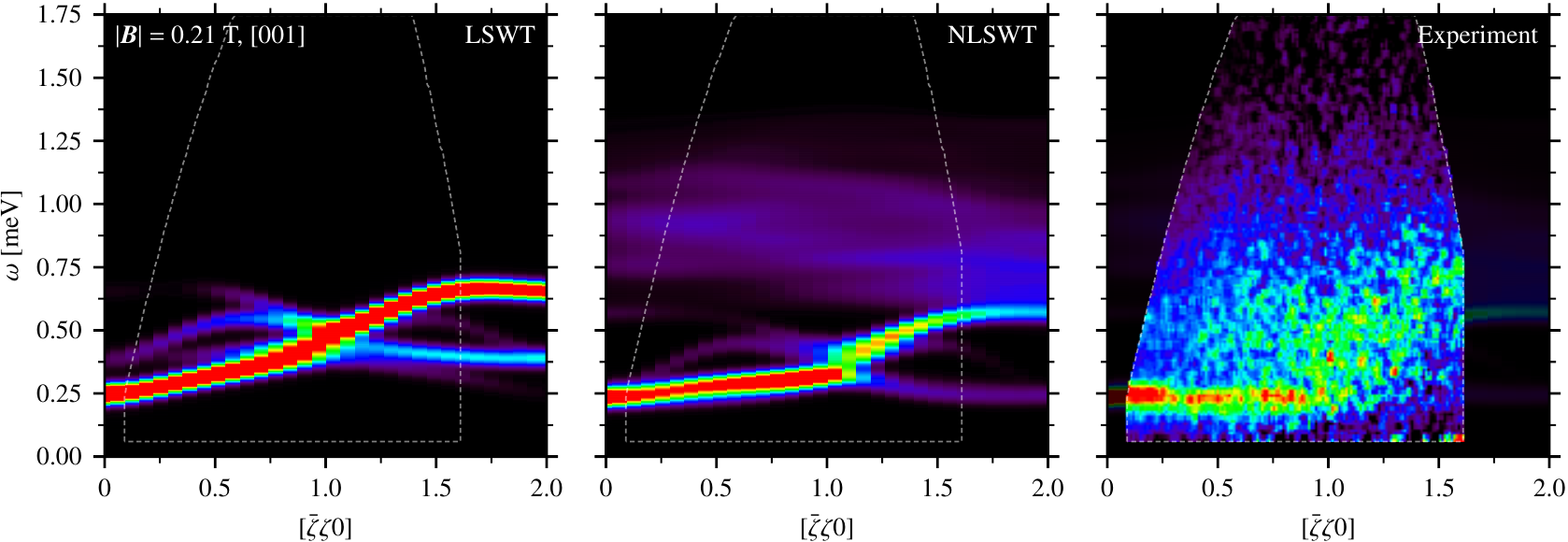}
    \put(-1,32.75){(c)}
  \end{overpic}\vspace{0.2cm}
  \begin{overpic}[width=0.85\textwidth]{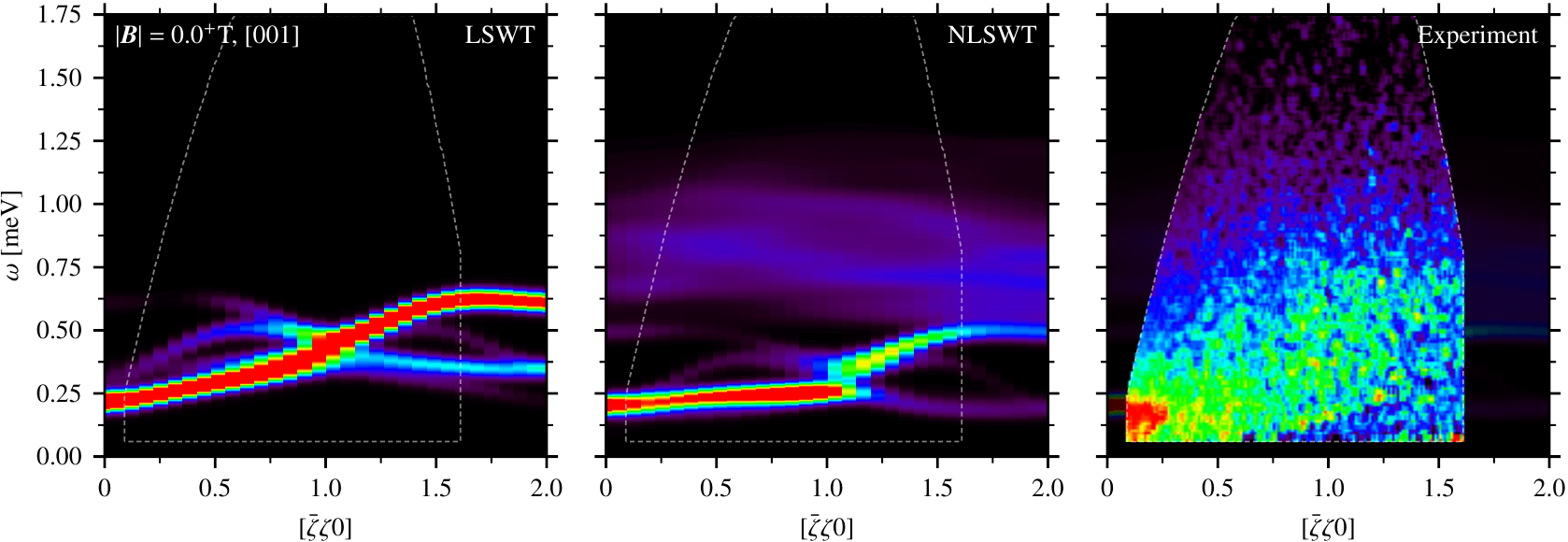}
    \put(-1,32.75){(d)}
  \end{overpic}
  \caption{\label{fig:exp1m10}
    (a-d) Comparison of inelastic neutron scattering intensity, $I(\vec{q},\omega)$ [Eq.~(\ref{eq:intensity})], in \acf{LSWT},  \acf{NLSWT} and in the experimental data of \citet{thompson2017} along $[\bar{\zeta}{\zeta}0]$ for four different fields along $[001]$, (a) $|\vec{B}| = 5\T$, (b) $1.5\T$, (c) $0.21\T$ and (d) $0.0^+\T$. Intensities in all plots are averaged  over wave-vectors along transverse directions over the range $l, k = [-0.2,0.2]$.  Overall scale is chosen to match experimental data, and is consistent across Figs.~\ref{fig:exp100}, \ref{fig:exp1m10}, and \ref{fig:exp110}. The dashed lines indicate the boundaries of the experimental data, while features outside these boundaries in the experimental panels correspond to the \ac{NLSWT} calculation. }
\end{figure*}

\begin{figure*}
  \begin{overpic}[width=0.85\textwidth]{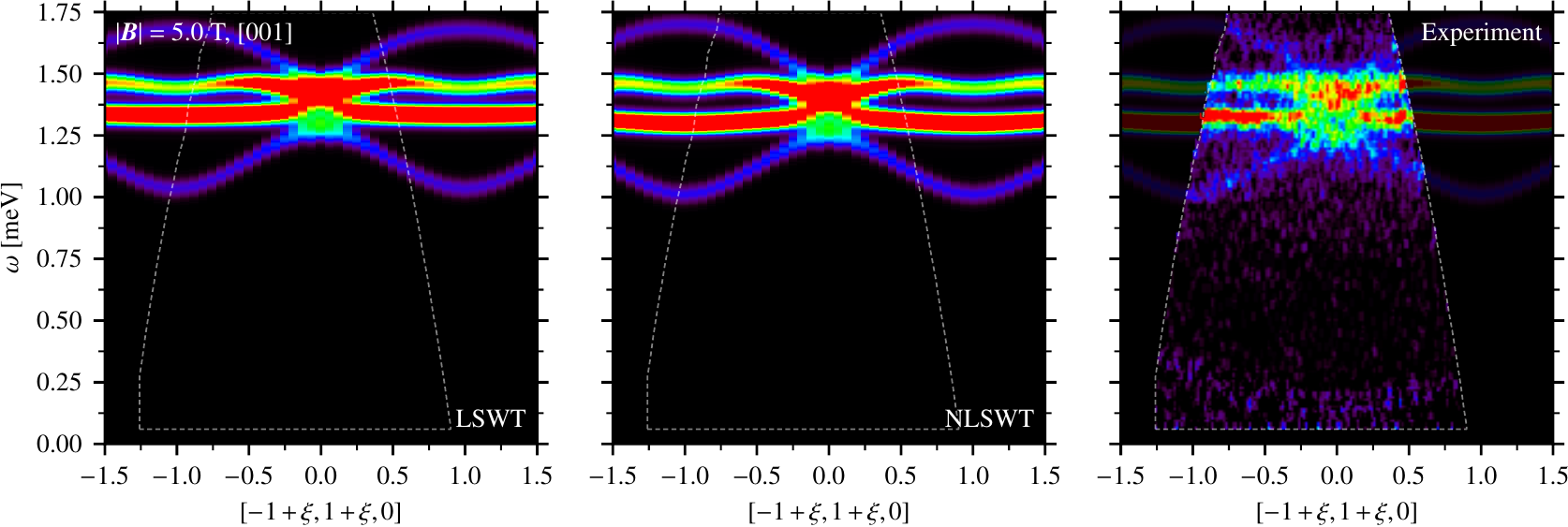}
    \put(-1,32){(a)}
  \end{overpic}\vspace{0.2cm}
  \begin{overpic}[width=0.85\textwidth]{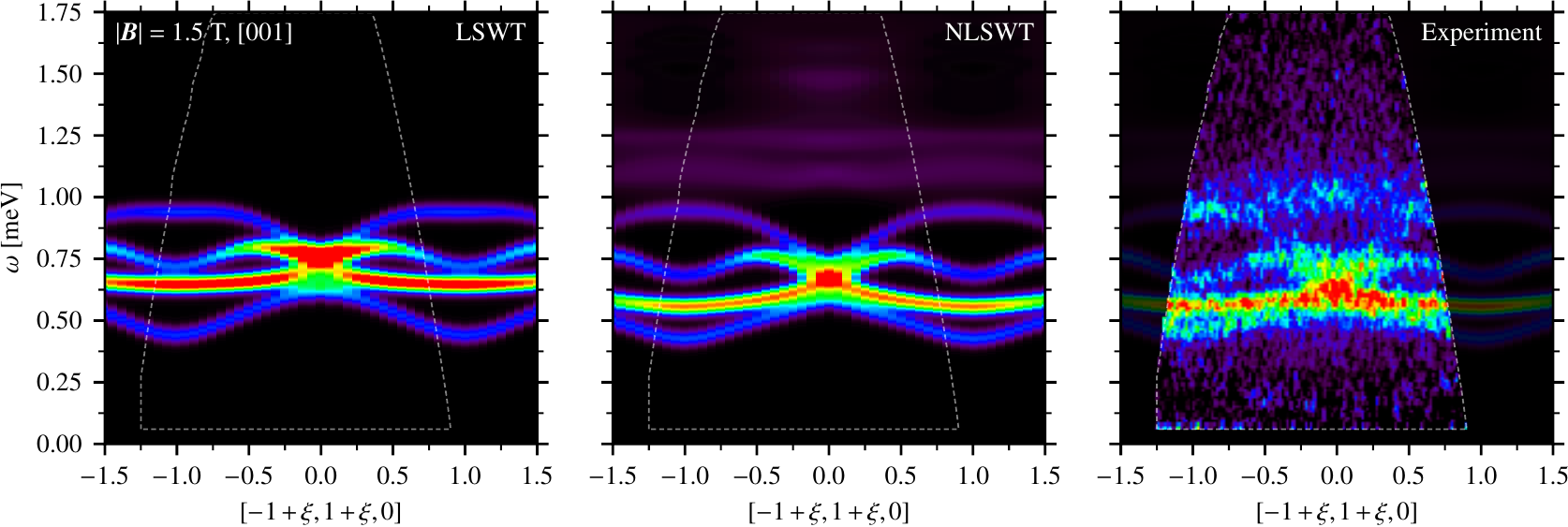}
    \put(-1,32){(b)}
  \end{overpic}\vspace{0.2cm}
  \begin{overpic}[width=0.85\textwidth]{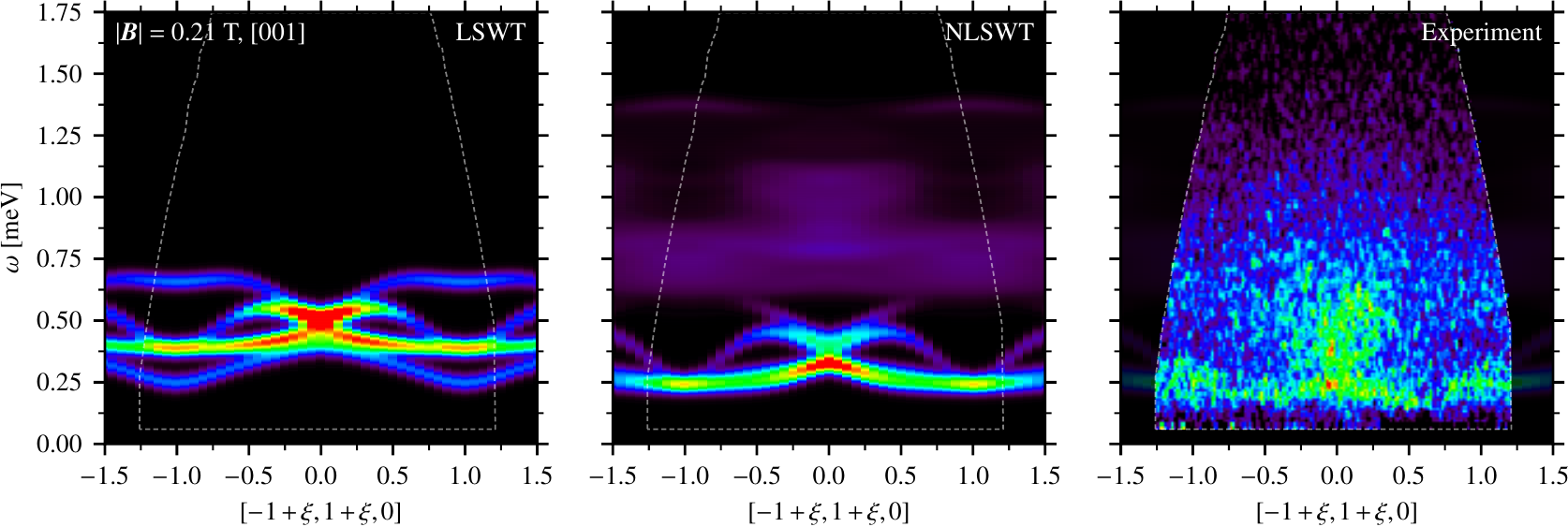}
    \put(-1,32){(c)}
  \end{overpic}\vspace{0.2cm}
  \begin{overpic}[width=0.85\textwidth]{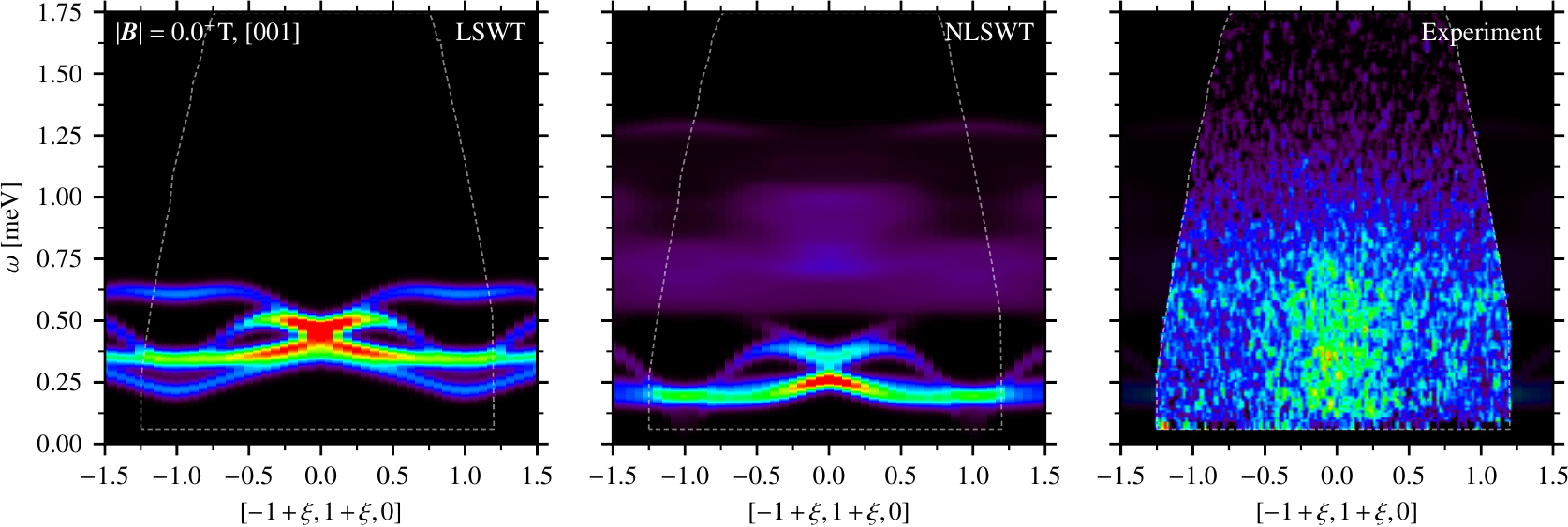}
    \put(-1,32){(d)}
  \end{overpic}  
  \caption{\label{fig:exp110}
    (a-d) Comparison of inelastic neutron scattering intensity, $I(\vec{q},\omega)$ [Eq.~(\ref{eq:intensity})], in \acf{LSWT},  \acf{NLSWT} and in the experimental data of Ref.~\cite{thompson2017} along $[-1+\xi,1+\xi,0]$ for four different fields along $[001]$, (a) $|\vec{B}| = 5\T$, (b) $1.5\T$, (c) $0.21\T$ and (d) $0.0^+\T$. Intensities in all plots are averaged over wave-vectors along transverse directions over the range $l, k = [-0.2,0.2]$.  Overall scale is chosen to match experimental data, and is consistent across Figs.~\ref{fig:exp100}, \ref{fig:exp1m10}, and \ref{fig:exp110}. The dashed lines indicate the boundaries of the experimental data, while features outside these boundaries in the experimental panels correspond to the \ac{NLSWT} calculation. }
\end{figure*}

We now consider a more direct comparison between the \ac{LSWT} and \ac{NLSWT} dynamical structure factor and the detailed experimental data in $[001]$ magnetic field, as studied in Ref.~\cite{thompson2017}. This field direction selects a single domain at low fields and adiabatically connects the low field ferromagnet to the polarized state. We focus on three cuts in momentum space: one along $[h00]$ (Fig.~\ref{fig:exp100}), one along $[\bar{\zeta}{\zeta}0]$ (Fig.~\ref{fig:exp1m10}) and one along $[-1+\xi,1+\xi,0]$ (Fig.~\ref{fig:exp110}). For each cut we integrate over a region out-of-plane and perpendicular to the cut line of width $[-0.2,0.2]$ r.l.u to allow direct comparison to experiment \cite{thompson2017}. We compute the full neutron scattering intensity, including the \rth{Yb} atomic form factor, as given in Eq.~(\ref{eq:intensity}). Throughout we use the parameters of Ref.~\cite{thompson2017} and fix the overall intensity scale globally, for all fields and cuts, to best match the $[h00]$ cut at $|\vec{B}|=5\T$. 

To frame our discussion, let us highlight three features of the data and calculations; first, we have the one-magnon-like features that are sharp in energy and wave-vector with high-intensity. Second, we have the extent of broadening of the magnon bands. Third, we have the intensity and location of the (primarily) two-magnon continuum. We note that, as the two-magnon continuum is not determined self-consistently in our \ac{NLSWT}, it will be systematically higher in energy than what may be expected from the one-magnon energies, which are renormalized downward.

As in the previous subsection, we find that the results at $|\vec{B}|=5\T$ match the experimental data very well, with \ac{NLSWT} offering no significant improvement over \ac{LSWT} [see Figs.~\subref{fig:exp100}{(a)}, \subref{fig:exp1m10}{(a)}, \subref{fig:exp110}{(a)}]. We note that the agreement of one-magnon energies from \ac{NLSWT} are slightly \emph{worse} than those from \ac{LSWT} [see, e.g., Fig.~\subref{fig:exp1m10}{(a)}]; this is to be expected given that these parameters were determined using \ac{LSWT} and there are small interaction corrections even at these high fields.   

At $|\vec{B}|=1.5\T$, the field is sufficiently low that the one- and two-magnon states now overlap. One finds qualitative agreement between \ac{NLSWT} and \ac{LSWT} for primarily one-magnon features, but with visible quantitative differences in the intensity of the two-magnon continuum, and in some broadening [e.g. near $[\bar{1}00]$ Fig.~\subref{fig:exp100}{(b)}]. Across each of the cuts these differences improve the agreement of \ac{NLSWT} with experiment relative to \ac{LSWT}. \ac{NLSWT} also includes some intensity from the two-magnon continuum, roughly in the same location as some of the broad, diffuse scattering seen experimentally [shown in Figs.~\subref{fig:exp100}{(b)}, \subref{fig:exp1m10}{(b)}, \subref{fig:exp110}{(b)}], though it is somewhat weaker than in the theoretical calculation.

At lower fields neither \ac{LSWT} nor \ac{NLSWT} give a full account of the experimental data. The experimental result includes a very diffuse background of scattering over a wide range in energies that is neither reproduced in \ac{NLSWT} by spontaneous magnon decay nor by the direct intensity of the two-magnon continuum. However, many of the one-magnon features of \ac{NLSWT} match well with the sharper features of the experimental data. For example, the nearly flat mode along $[h00]$ at $|\vec{B}|=0.21\T$ [see Fig.~\subref{fig:exp100}{(c)}] that is observed experimentally is also seen in \ac{NLSWT}, but not in \ac{LSWT} where it has significantly more dispersion. We find similar rough agreement also for the location of the gross features of the $[\bar{\zeta}{\zeta}0]$ and $[-1+\xi,1+\xi,0]$ cuts [see Figs.~\subref{fig:exp1m10}{(c)}, \subref{fig:exp110}{(c)}]. As in the $1.5\T$ data the intensity of the two-magnon continuum is also markedly lower in \ac{NLSWT} than in the experimental data. In the zero-field result, this trend is enhanced, with the experimental data becoming broader and more diffuse, but with the gross intensity features being in roughly the same locations as in \ac{NLSWT} [see Figs.~\subref{fig:exp100}{(d)}, \subref{fig:exp1m10}{(d)}, \subref{fig:exp110}{(d)}].

\subsection{Powder-averaged inelastic neutron scattering}
\label{sec:powder}

\begin{figure*}[htp]
  \centering
  \includegraphics[width=0.75\textwidth]{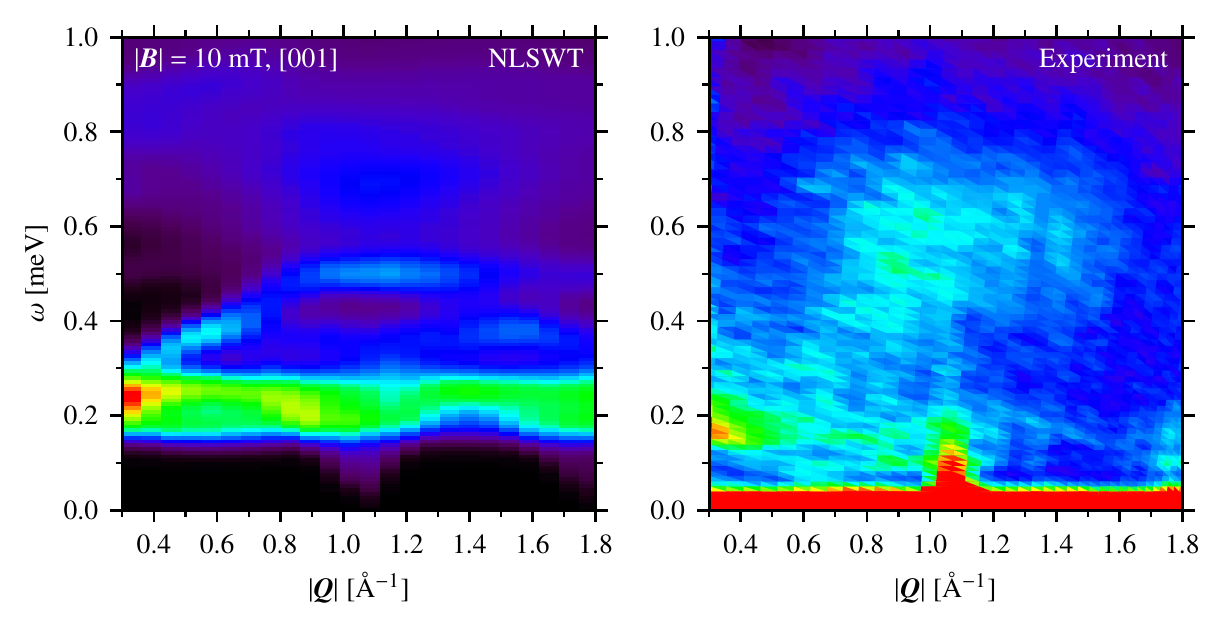}
  \caption{\label{fig:powder} 
    Comparison of \acf{NLSWT} calculation of powder-averaged inelastic neutron scattering intensity [see Eq.~(\ref{eq:intensity})] and zero-field experimental data of \citet{VPA2017}. A small field has been applied in the theoretical calculation to resolve the zero field instability for the parameters of Ref.~\cite{thompson2017} (see Sec.~\ref{sec:critical-fields}), as well as some broadening to mimic the finite experimental energy resolution.
  }
\end{figure*}

Given the sample dependence present in crystals of \yto{} grown under different conditions or by different groups, it is worthwhile to make some comparisons to other data sets.  Here we consider the inelastic data on the powder of~\citet{VPA2017}, as it has higher energy resolution and shows more structure than (similar) powder data obtained by other groups. Other powder samples, for example as presented in the review of~\citet{HallasReview}, show only a broad continuum, similar to what is seen in the single-crystal of \citet{thompson2017}. Since we have compared to this data at length in Sec.~\ref{sec:direct-comparison}, we do not make direct comparisons to these other powder samples here.

For a zero-field cooled sample, powder-averaging can be carried out straightforwardly, by integrating over wave-vectors with a given magnitude.  We present the powder averaged intensity [Eq.~(\ref{eq:intensity})] for the parameters of Ref.~\cite{thompson2017} in Fig.~\ref{fig:powder}. We have added a small ($10\mT$) $[001]$ field to render the \ac{SFM} state stable. For comparison we also include the data of \citet{VPA2017} presented similarly. We see that several of the gross features are reproduced: a gap and band of intensity near $\sim 0.2\meV$ which is maximal at small $|\vec{Q}|$ and a broad continuum of intensity extending from $\sim 0.3\meV$ out to $\sim 0.8\meV$ with maximal intensity near $\sim 1\AA^{-1}$. Several features are absent however, most notably the high intensity near $1.1\AA^{-1}$ and the sub-gap features visible in the range $0.5\AA^{-1} \lesssim |\vec{Q}| \lesssim 1.1\AA^{-1}$. We do note that this intensity near $1.1\AA^{-1}$ coincides with location of the nearly gapless mode at $[111]$ that is present in our calculation. However the intensity of this mode does not match; as the minimum only occurs in a small region of momentum space, only a weak intensity is visible extending down at $1.1\AA^{-1}$ in the \ac{NLSWT} calculation. As in Sec.~\ref{sec:direct-comparison}, the intensity of the higher energy two-magnon continuum is underestimated by the \ac{NLSWT}.

\section{Discussion}
\label{sec:disc}

We have presented a detailed study of \ac{NLSWT} as applied to \yto. In the fully polarized phase where fluctuations are small and multi-magnon continua are well-separated from the one-magnon states, \ac{LSWT} is expected to provide a quantitative description of the one-magnon dispersions. This observation motivated fits of \ac{LSWT} to inelastic neutron scattering data at $5\T$ and above where the magnetic field energy is comfortably greater than the exchange scale. Our \ac{NLSWT} calculations confirm that the effects of magnon interactions are weak at these fields providing a retrospective justification for the \ac{LSWT} fits. 

The corrections from \ac{NLSWT} are expected to be more significant at lower fields, and to be quantitative within some window of intermediate fields. Indeed, we find that the critical fields predicted by \ac{LSWT} are significantly renormalized by magnon interactions in the direction expected on the basis of experiments (Sec.~\ref{sec:critical-fields}). By comparing with inelastic neutron scattering data (Sec.~\ref{sec:DSF}), we have found empirically that one-magnon energy renormalizations and spontaneous magnon decay within the higher energy magnon bands starts to become significant for $[001]$ and $[1\bar{1}0]$ fields $\lesssim 2 \T$ and for $[111]$ fields $\lesssim 3 \T$. Comparison to Ref.~\cite{thompson2017}, which contains energy-momentum slices for various $[001]$ fields, suggests that \ac{NLSWT} is nearly quantitative at $\sim 1.5\T$, and captures some aspect of the experimental intensity reasonably well down to even $\sim 0.2 \T$ (see Figs.~\ref{fig:exp100}, \ref{fig:exp1m10} and \ref{fig:exp110}), but misses the extensive broadening. This is a central result of this paper.

We turn now to the low field data. Inelastic neutron scattering data at zero field can be found as powder averages in Refs.~\cite{gaudetgapless2016,VPA2017} and in single crystal samples in Refs.~\cite{Robert2015,thompson2017}. In Refs.~\cite{Robert2015,gaudetgapless2016,thompson2017} the scattering intensity appears to have no sharp features within experimental resolution, though clear intensity modulations are seen within the apparent continuum; see  Figs.~\subref{fig:exp100}{(d)}, \subref{fig:exp1m10}{(d)} and \subref{fig:exp110}{(d)} for reproductions of the Ref.~\cite{thompson2017} zero field data.  The powder data of \citet{VPA2017} is compared with the ($10 \mT$) \ac{NLSWT} calculation using the parameters of \citet{thompson2017}. While continuum scattering is visible in that data above $\sim 0.3 \meV$, the low energy sharp features present in the calculation,  such as the flat band of intensity at about $\sim 0.2 \meV$, are present in the data. 

In the calculation, there is an additional soft mode with intensity at $[111]$. We note that there is a corresponding feature in the data of \citet{VPA2017}, albeit with a higher intensity than predicted by \ac{NLSWT}.  While some evidence has been presented~\cite{changhfm2016} for a $\sim 40 \mueV$ soft mode at $[111]$ in a single crystal via inelastic neutron scattering at low fields, this has not been explored in detail in the literature. That this gap has not been seen in other crystals may simply be a consequence of the smallness of the gap relative to the energy resolution of these other experiments.  

On the theoretical side, the presence of such a soft mode is expected near the phase boundary between the \ac{SFM} ground state of \yto{} and the antiferromagnetic $\Gamma_5$ manifold~\cite{Jaubert2015,yan2017}, as has been argued to be relevant for the exchange parameters of Refs.~\cite{ross2011,Robert2015,thompson2017}. The phase boundary obtained classically is renormalized by quantum fluctuations~\cite{Jaubert2015}, and we find that the parameters of Ref.~\cite{thompson2017}, noted as being close to the classical phase boundary, are even closer to the renormalized instability line. These observations suggest that \yto{} may be closer to the \ac{SFM}-$\Gamma_5$ phase boundary than would be expected on the basis of a classical calculation from the empirically determined exchange parameters.

Furthermore, given the presence of a gapped mode at the zone center, one can make the model-independent observation that na\"ive kinematics already forbid broadening of this mode via magnon interactions. Leading order \ac{NLSWT}, as we have presented here, has difficulty capturing this quantitatively, as the two-magnon continuum is not determined self-consistently and reflects the bare one-magnon kinematics. However, if a self-consistent treatment that resolves such issues were carried out, it would likely not be able to capture the decay region seen experimentally, given it still lies outside the (loose) bounds defined by even the \emph{experimental} one-magnon features. These conclusions are unaffected by the presence or absence of the nearly soft mode at $[111]$. While such a soft mode can place much of the rest of the one-magnon spectrum in the two-magnon continuum, the density of states associated with this mode is too small to induce any significant linewidth. One potential resolution could be in a breakdown of the na\"ive kinematics described above through strong interactions within the two-magnon continuum~\cite{zhitomirsky2013colloquium,verresen2018strong}.

\section{Outlook}
\label{sec:outlook}

We have shown that \ac{NLSWT} describes the single crystal inelastic neutron scattering data at intermediate strength $[001]$ magnetic fields, capturing some of the observed broadening of the one-magnon lineshapes and the renormalization of the one-magnon energies. At lower fields, the theory fails to account for the anomalously broadened magnetic scattering observed in various powder and single crystal samples. There are several possible scenarios for how this could be resolved, which we loosely categorize as intrinsic and extrinsic, i.e. based on a more detailed analysis of the present model, or involving ingredients beyond it.

Several possible intrinsic effects, beyond the magnon interactions we have considered here, could 
play a role in the low-field physics of \yto{}. We have described ways in which we expect our calculations to fall short of capturing the full extent, at low fields, of the interaction corrections to \ac{LSWT}. Further investigation of the renormalization of the two-magnon states and their self-consistent influence on the one-magnon branches may be useful in this regard. We stress, however, that for this kind of explanation to capture the zero field excitations in \yto{}, the na\"ive kinematics outlined in Sec.~\ref{sec:disc} must break down to allow the broadening to persist to the lowest energies, as observed experimentally.

Also, it is possible that multi-magnon effects involving three or more magnons may be required to understand the anomalous broadening. In such a case it may make sense to describe these excitations, while still arising from the conventional ferromagnetic phase, as being related to some nearby exotic ``parent'' state, such as some kind of quantum spin liquid~\cite{ross2011,Chang2012,sanyal2018interplay,chern2018magnetic}. Similar scenarios have been outlined to understand the high-energy excitations in the Kitaev magnet $\alpha$-RuCl\tsub{3}, with the Kitaev spin liquid serving as the parent state~\cite{banerjee2016proximate}. Given the proximity to the \ac{SFM}-$\Gamma_5$ phase boundary, it is not implausible that quantum effects could stabilize a disordered intermediate phase near the boundary~\cite{sanyal2018interplay,chern2018magnetic} that could serve as parent state.

While intrinsic effects are clearly important at low fields, the strong sample to sample variations that have been reported~\cite{Ross2012a,gaudet2015,Mostaed2017,shafieizadeh2018superdislocations,bowman2019role} in \yto{}, as well as the issues raised in Sec.~\ref{sec:disc}, suggests that extrinsic effects may also be important at low fields. This calls into question whether further refinements to spin wave theory are capable of explaining all of the puzzling aspects of the zero field data. This sample dependence, along with the sensitivity to applied hydrostatic~\cite{kermarrec2017ground} or chemical~\cite{HallasReview} pressure, further supports an extrinsic origin of some of the broadening exhibited at low fields. If, indeed, extrinsic effects contribute significantly to the scattering continuum, sufficiently clean crystals may exhibit inelastic spectra more in line with predictions of \ac{NLSWT}. From this perspective, the relatively sharp features in the zero field inelastic powder data of \cite{VPA2017} and the sharp low energy soft mode reported in \cite{changhfm2016} might be interpreted as arising from the underlying intrinsic physics. 

Extrinsic effects, such as structural disorder, can take many distinct forms. Several different types of structural disorder have been discussed for \yto{}: these include point-defects, such as ``stuffed'' \rth{Yb} spins~\cite{Ross2012a,gaudet2015,bowman2019role}, oxygen vacancies~\cite{Mostaed2017,bowman2019role}, as well as internal strains and atomic displacements, such as those that might arise due to line-defects like dislocations~\cite{shafieizadeh2018superdislocations} or through intrinsic frustration of the structural network~\cite{trump2018universal}. For example, recent experiments have explored the effects of oxygen annealing and stuffed \rth{Yb} spins~\cite{bowman2019role}, finding that the presence or absence oxygen vacancies has strong effects on the magnetic correlations. It has also been reported~\cite{shafieizadeh2018superdislocations} that some single crystals are liable to contain defects such as ``superdislocations'' or anti-phase boundaries that can induce further lattice distortions. 

Any of these defects could provide a route to broadening the spectrum. However, given the (nominally) low-level of such defects reported in many samples~\cite{bowman2019role,Arpino2017},  one then must grapple with the question of how (nominally) small amounts of disorder could lead to the large observed broadening. Indeed, the sensitivity~\cite{Arpino2017} to such extrinsic effects suggests that the physics of \yto{} may be quite delicate. One potential route to rendering the magnetic physics delicate, and thus extremely sensitive to disorder, is close proximity to the \ac{SFM}-$\Gamma_5$ boundary~\cite{Jaubert2015,Robert2015}. One could then  imagine that defects disturb the local physics enough to mix the \ac{SFM} and $\Gamma_5$ states, strongly affecting the nature of the dynamical response. 

As a concrete example, one can consider domain walls between the different \ac{SFM} domains. Near the boundary between the \ac{SFM} and $\Gamma_5$ phases, these domain walls in fact (locally) pass through the manifold of $\Gamma_5$ states, reflecting the topology of the piecewise connected, (nearly) degenerate $U(1)$ manifolds~\cite{Canals2008,Chern2010,yan2017}. The width of these domain walls diverges (classically) as one approaches the boundary, potentially providing a route to significant broadening even for relatively low levels of disorder.~\footnote{We note that such a possibility, with the $\Gamma_5$ states appearing in \ac{SFM} domain walls, was recently independently noted by the Johns Hopkins group~\cite{scheietalk}}

Crucially, as we showed in Sec.~\ref{sec:phase}, the effects of magnon interactions appear to move \yto{} even closer to the phase boundary, further enhancing any such effects. Such a scenario would also be largely compatible with the dynamics being simpler at high-fields: once the magnetic field energy lifting the near-degeneracy at low fields dominates the disorder energy scales, the phase would simply be a \ac{SFM} with a small amount of disorder, and thus would likely exhibit a significantly smaller amount of broadening than when the \ac{SFM} and $\Gamma_5$ states are mixed. 
With both intrinsic and extrinsic scenarios in contention, there remains a great deal to do to reach a quantitative understanding of the zero field spectrum in \yto{}. Given the progress in understanding obtained starting from the high field limit, we are optimistic about the prospects for settling this question in \yto{} in the near future.

The importance of resolving this issue is highlighted by the the existence of puzzling continua in other magnetic materials. While in the foregoing discussion we have focused on the experimental results in \yto{}, there are tantalizing results~\cite{HallasReview} that suggest that the physics seen in \yto{} also appears in some other, less explored, materials such as \yso{} and \ygo{}. These two compounds are isostructural to \yto{}, with the substitution of the non-magnetic Ti for Sn or Ge expected mainly to act as chemical pressure~\cite{dunPressure}. One, \yso{}, is known to have a \ac{SFM} ground state~\cite{dun2013yb}. while \ygo{}~\cite{hallas2016xy} shows $\Gamma_5$ antiferromagnetic order. However, \emph{both} show a similar broad continuum of excitations~\cite{hallas2016xy,HallasReview} at zero-field. While a detailed parametrization of the exchange in \yso{} or \ygo{} has not yet been carried out, it is plausible that they also live close to the \ac{SFM}-$\Gamma_5$ phase boundary~\cite{Jaubert2015}, like \yto{}. In the same vein, it has been argued~\cite{reotier2017} that the rare-earth spinels AYb\tsub{2}X\tsub{4} (A=Cd, Mg and X=S,Se)~\cite{lau2005,higo2017,reotier2017}, which exhibit $\Gamma_5$ order~\cite{reotier2017}, could share much of the physics. However, given their different crystal structure, extrinsic effects could manifest themselves differently and may shed some light on the physics of \yto{}, \ygo{} and \yso{}.

\begin{acknowledgments}
We thank R. Coldea and M. Gingras for helpful comments and collaborations on prior work. 
We also thank  R. Coldea as well as V. Pe\c{c}anha-Antonio and Y. Su for providing experimental data (Refs.~\cite{thompson2017} and \cite{VPA2017} respectively). This work was in part supported by Deutsche Forschungsgemeinschaft (DFG) under grant SFB 1143 and through the W\"urzburg-Dresden Cluster of Excellence on Complexity and Topology in Quantum Matter -- \textit{ct.qmat} (EXC 2147, project-id {390}{854}{90}).
\end{acknowledgments}

\appendix

\section{Conventions}
\label{app:conventions}

We choose the four local frames $(\vhat{x}_i,\vhat{y}_i,\vhat{z}_i)$ for the pyrochlore lattice following \citet{savary2012}
\begin{align}
  \vhat{z}_1 &= \frac{1}{\sqrt{3}} \left(+\vhat{x}+\vhat{y}+\vhat{z}\right), &
  \vhat{x}_1 &= \frac{1}{\sqrt{6}} \left(-2\vhat{x}+\vhat{y}+\vhat{z}\right),   \nonumber
  \\
  \vhat{z}_2 &= \frac{1}{\sqrt{3}} \left(+\vhat{x}-\vhat{y}-\vhat{z}\right), &
  \vhat{x}_2 &= \frac{1}{\sqrt{6}} \left(-2\vhat{x}-\vhat{y}-\vhat{z}\right),   \nonumber
  \\
  \vhat{z}_3 &= \frac{1}{\sqrt{3}} \left(-\vhat{x}+\vhat{y}-\vhat{z}\right), &
  \vhat{x}_3 &= \frac{1}{\sqrt{6}} \left(+2\vhat{x}+\vhat{y}-\vhat{z}\right),   \nonumber
  \\  \vhat{z}_4 &= \frac{1}{\sqrt{3}} \left(-\vhat{x}-\vhat{y}+\vhat{z}\right), &
  \vhat{x}_4 &= \frac{1}{\sqrt{6}} \left(+2\vhat{x}-\vhat{y}+\vhat{z}\right),
\end{align}
where $\vhat{y}_i = \vhat{z}_i \times \vhat{x}_i$. The four basis sites of the tetrahedron are then along the $\vhat{z}_i$ directions, with $\vhat{r}_i = \vhat{z}_i$.

Exchange parameters have been reported in the main text in this local frame, via $J_{zz}$, $J_{\pm}$, $J_{\pm\pm}$ and $J_{z\pm}$ and in the dual global frame~\cite{PhysRevB.98.054408} where the symmetry-allowed couplings on one pair of nearest neighbors (sublattices $1$ and $2$) can be written
\begin{equation}
\left( \begin{array}{ccc} 
J+K & +D/\sqrt{2} & +D/\sqrt{2} \\   
-D/\sqrt{2} & J & \Gamma \\
-D/\sqrt{2} & \Gamma & J
\end{array}  \right),
\end{equation}
with transformation
\begin{equation}
\left( \begin{array}{c} 
J \\ K \\ \Gamma \\ D
\end{array}  \right) =
\frac{1}{3}\left( \begin{array}{cccc} 
                    -1 & +4 & +2 & -2\sqrt{2} \\
                    +2 & -8 & +2 & -2\sqrt{2} \\
                    -1 & -2 & -4 & -2\sqrt{2} \\
                    -\sqrt{2} & -2\sqrt{2} & +2\sqrt{2} & +2
\end{array}  \right)
\left( \begin{array}{c} 
J_{zz} \\ J_{\pm} \\ J_{\pm\pm} \\ J_{z\pm}
       \end{array}  \right),
\end{equation}
between the exchange couplings. Here $J$, $K$, $D$, $\Gamma$ are respectively couplings for (dual) Heisenberg, Kitaev Dzyaloshinskii-Moriya, and symmetric off-diagonal exchange couplings.

\bibliography{draft}

\end{document}